# Blockchain-based Recommender Systems: Applications, Challenges and Future Opportunities

Yassine Himeur[a,i], Aya Sayed[a], Abdullah Alsalemi[a,b], Faycal Bensaali[a], Abbes Amira[c,b], Iraklis Varlamis[d], Magdalini Eirinaki[e], Christos Sardianos[d] and George Dimitrakopoulos[d]

[a]*Department of Electrical Engineering Qatar University Doha Qatar*
[b]*Institute of Artificial Intelligence De Montfort University Leicester United Kingdom*
[c]*Department of Computer Science University of Sharjah UAE*
[d]*Dept. of Informatics and Telematics Harokopio University of Athens Greece*
[e]*San Jose State University San Jose CA USA*

ABSTRACT

Recommender systems have been widely used in different application domains including energy-preservation, e-commerce, healthcare, social media, etc. Such applications require the analysis and mining of massive amounts of various types of user data, including demographics, preferences, social interactions, etc. in order to develop accurate and precise recommender systems. Such datasets often include sensitive information, yet most recommender systems are focusing on the models' accuracy and ignore issues related to security and the users' privacy. Despite the efforts to overcome these problems using different risk reduction techniques, none of them has been completely successful in ensuring cryptographic security and protection of the users' private information. To bridge this gap, the blockchain technology is presented as a promising strategy to promote security and privacy preservation in recommender systems, not only because of its security and privacy salient features, but also due to its resilience, adaptability, fault tolerance and trust characteristics. This paper presents a holistic review of blockchain-based recommender systems covering challenges, open issues and solutions. Accordingly, a well-designed taxonomy is introduced to describe the security and privacy challenges, overview existing frameworks and discuss their applications and benefits when using blockchain before indicating opportunities for future research.

## 1. Introduction

Recommender systems (RSs), initially introduced to address the problem of improving the customer experience and retention in e-commerce sites Schafer et al. (1999) has since become a ubiquitous and often anticipated functionality of many online interactions, from movie and song recommendations Deldjoo et al. (2020b) to applications related to tourism Hong and Jung (2021), social networks Ghafari et al. (2020), health Saha et al. (2020), energy and smart cities Quijano-Sánchez et al. (2020), and many more.

The rise in popularity of Internet of things (IoT), Internet-connected devices (e.g. mobile phones, tablets, laptops, smart sensors, etc.), and sensors, has resulted in the generation of massive amounts of data Deebak and Al-Turjman (2020); Katarya and Verma (2017). These huge quantities of data, characterized by the "five V's", namely *velocity*, *variety*, *veracity*, *volume*, and *value* is referred to as *big data* Gupta and Katarya (2018). Having the ability to both generate and process big data dramatically transformed many aspects of everyday life, including social network interactions, e-commerce, healthcare services, energy etc Katarya and Verma (2018). Appropriate processing of big data can provide the users with significant knowledge about their health, environment, and others Fu et al. (2018) and allows them to adapt to changes on time. However, as pervasive and ubiquitous such applications are, the characteristics of big data makes processing them a challenging problem.

While big data analytics have long been used to support humans in processing such input and making decisions, the need to introduce new, scalable RS algorithms that go beyond the simple rating-based input and can incorporate diverse types and huge amounts of data has become evident Zhang et al. (2019). In this context, the significant advancement of machine learning has helped the RS community to move from traditional RSs, that use clustering, nearest neighbors, matrix factorization, and collaborative filtering Gupta and Katarya (2019); Katarya (2018), to a new generation of RSs, which are powered by complex deep learning systems Gupta and Katarya (2021a) and knowledge graph Guo et al. (2020). This has brought rise to a broad and diverse set of application domains for RSs, including e-commerce Jerripothula et al. (2020), social media Anandhan et al. (2018) and networks Eirinaki et al. (2018), e-learning Wu et al. (2015), social behavior analysis Chen et al. (2019), energy saving Himeur et al. (2021), healthcare Deng and Huangfu (2019), IoT Wei et al. (2020), tourism Borràs et al. (2014), fashion Nguyen et al. (2014), and food industry Toledo et al. (2019), among others. Table 1 presents a summary of various data sources that are used as input in such RSs, in addition to the typical *<user, item, rating>*

yassine.himeur@qu.edu.qa (Y. Himeur); as1516645@qu.edu.qa (A. Sayed); a.alsalemi@qu.edu.qa,abdullah.alsalemi@my365.dmu.ac.uk (A. Alsalemi); f.bensaali@qu.edu.qa (F. Bensaali);
aamira@sharjah.ac.aea,abbes.amira@dmu.ac.uk (A. Amira);
varlamis@hua.gr (I. Varlamis); magdalini.eirinaki@sjsu.edu (M. Eirinaki);
sardianos@hua.gr (C. Sardianos); gdimitra@hua.gr (G. Dimitrakopoulos)
ORCID(s):

**Table 1**
Summary of the types of information used by Recommendation Systems.

| Information | Description |
| --- | --- |
| Item attributes | Descriptive information about the items (i.e. their features). Examples include brand, color, model, category, place of origin, etc. |
| User attributes | Descriptive information about the users (i.e. their features). Examples include age, marital status, education, demographics etc. |
| User ratings for the items | Explicit user feedback, in the form of ratings. can be scalar or binary. |
| Implicit user preferences | Information that is implicitly derived and relates to the user's choices. Examples are clicks, tags, and comments. |
| Recommendation feedback | The user response to the recommendations. It is expressed as accept/reject values, positive or negative labels, etc. Can be used to define (implicitly and explicitly) the user preferences. |
| User behavioural information | Implicit data recorded during the interaction of the user with the broader system. |
| Contextual information | Information on the context of recommendations. Examples are time, date, location, user status etc. |
| Social information | Data related to the user's social graph, including connections and interactions with other users, friendship relations (or similar) to other users, community membership, or both. |
| Domain knowledge | Background or prior information, empirical knowledge and rules that define the relation between content items and the user stereotype. This type of knowledge is usually static, but can also vary over time. |
| User purchase or consumption history | List of content items that have previously been purchased or consumed by the user. |

triplets, demonstrating the diversity of information used in RSs.

The value and sensitivity of such user-related data is undeniable. While several researchers have looked into developing privacy-preserving RSs WAN (2018), the majority of research that concerns developing new algorithms and models disregards this very important aspect of privacy and security, focusing on optimizing for accuracy and/or scalability. In addition, RSs are vulnerable to external attacks (e.g. injecting or manipulating data used for training the models) by adversaries. When focusing on security and privacy, ensuring that the data is safely stored has become a much more challenging task in the age of cloud (and thus decentralized) computing. Therefore, while there have been efforts to employ various risk reduction techniques, none of them has been completely successful, especially with regards to cryptographic security and the protection of the users' private information. In the past few years, we have observed the success of the blockchain technology as a decentralized way to ensure security and privacy preservation. This technology can be leveraged to solve one of the biggest challenges in RSs and thus this cross-domain area has started to gain interest from researchers in both communities.

In that context, by integrating blockchain into RSs, it becomes possible to (i) build trust-based and secure systems using the benefits of blockchain-supported secure multi-party computation, that adds smart contracts to the main blockchain based RS protocol; (ii) secure the control of users' information because the blockchain enables a safe data processing for the users in online portals. This is mainly due to its secure distributed ledger used to store data transactions; and (iii) provide the essential mechanisms for ensuring data protection and data integrity by adopting some encryption technologies. All these features together, along with the complicated structure and algorithms of blockchain, make blockchain-based RSs harder to manipulate and tamper with in comparison with conventional storage systems.

To that end, this paper presents, to the best of the authors' knowledge, the first review of blockchain-based RSs. We adopt a well-defined taxonomy to categorize the state-of-the-art blockchain-based RSs We first discuss the security and privacy challenges in RSs along with the characteristics, working concept, and the blockchain's potential impact on RSs. Subsequently, the blockchain-based RS architecture is explained followed by an overview of the main types of blockchain-based RSs. Next, the applications of blockchain-based RSs are described and various examples are portrayed. We then conduct a critical analysis to identify the drawbacks and limitations as well as open issues. Finally, we discuss relevant future research directions that will attract significant research and development in the near and far future. To summarize, the contributions of this work are as follows:

- It develops a bridge between two research worlds that can significantly benefit from each other, namely "*blockchain*" and "*recommender systems (RSs)*". It creates a unique map for researchers on blockchain-based RSs that comprises challenges, open issues, and solutions and illustrates the main application domains.

- It provides a framework for the development of decentralized RSs that take advantage of the blockchain

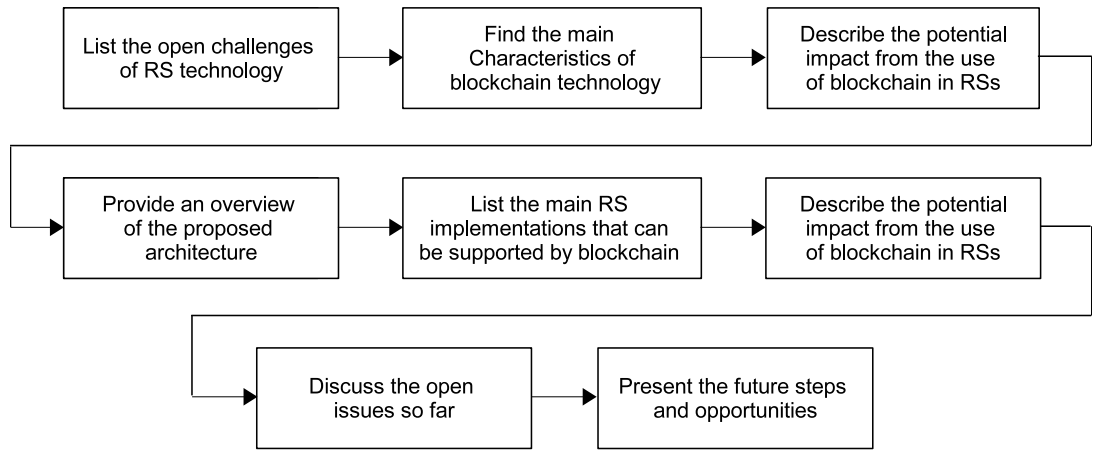

**Figure 1**: The main steps for introducing blockchain-based RS solutions and going beyond the current state of the art.

ledger to securely store data and separate them from services.

- It compares the performance of various existing frameworks with respect to different parameters, including time computation and accuracy of recommendations, to inform the state-of-the-art.

- It examines horizontal issues of RSs, such as scalability, explainability, distributed processing, and load balancing, which raise new research challenges for blockchain researchers and developers.

- It provides a set of future directions and orientations for improving the performance of blockchain-based RSs and addressing some of the current challenges.

This paper follows a specific workflow in order to cover the novel field of blockchain-based RSs and pave the path for researchers in the field, as depicted in Fig. 1. We begin with the open challenges of the more mature technology, which is RSs, and the needs that blockchain can cover. We continue with the main features of blockchain technology and explain how they can fit the requirements of RSs. We then proposes a core architecture for blockchain-based RSs and details on the main implementation types that are available. The next steps comprise the overview of application domains for the new solution and the open issues that have arisen so far. Finally, the future research opportunities are presented.

The rest of the paper is organized as follows: Section 2 provides an overview of blockchain-based RSs, starting with the main challenges of RSs and the main blockchain features that can be of benefit to RSs, followed by a discussion on the impact of blockchain's use to future RSs, a presentation of the main architecture of a blockchain-based RS, and the various types of blockchain-based RSs and their applications. Section 3 includes a critical analysis that highlights the drawbacks and limitations of the proposed technology and discusses the main open issues, whereas Section 4 provides the directions for future research on the topic. Finally, Section 5 concludes the paper by highlighting the importance of the proposed framework for the blockchain and RSs research community.

## 2. Overview of blockchain-based RSs

The use of blockchain technology in RSs is rather new, with the first paper been published in 2016 Frey et al. (2016a), and the majority of related work having been published within the last two years. In order to provide a proper coverage of the field and discover the challenges and opportunities for further research, it is important: i) to perform a review of related works and discuss their contributions, and ii) to organize these into a comprehensive taxonomy that will highlight the main research sub-areas and the active research topics in each one of them.

This will allow readers to get a better understanding of the various application domains of blockchain-based RSs, learn about the characteristics of the two domains (blockchain and RSs) and become aware of the challenges that arise from their merging. At the same time, it will allow researchers to gain useful insights on the future research directions on this field and beyond, especially in emerging technologies that are expected to affect research in blockchain-based RSs in the near future. In order to better organise the survey of related work in the area of blockchain-based RSs, we follow the taxonomy illustrated in Fig. 2.

### 2.1. Security and privacy challenges in RSs (C)

As in all machine learning algorithms, the success of RS models relies on the quality and quantity of data. The more the system knows about the past history of a user, the better the quality of the recommendations. Moreover, when additional information such as the one included in Table 1 is incorporated, and the more accurate and detailed it is, the more useful and personalized the recommendations made to the user are. In this context, due to the sensitivity of users' data, various security and privacy issues emerge from an illegal access or vulnerable processing of such data Du et al. (2018). The more specific the information about user profile, preferences, habits and actions, the larger the potential of leaking and threatening the user's privacy. Unfortunately, current RS algorithms and implementations usually neglect the potential leaks of personally identifiable information that risk of malicious attacks and the threats they can bring to user privacy Deldjoo et al. (2021).

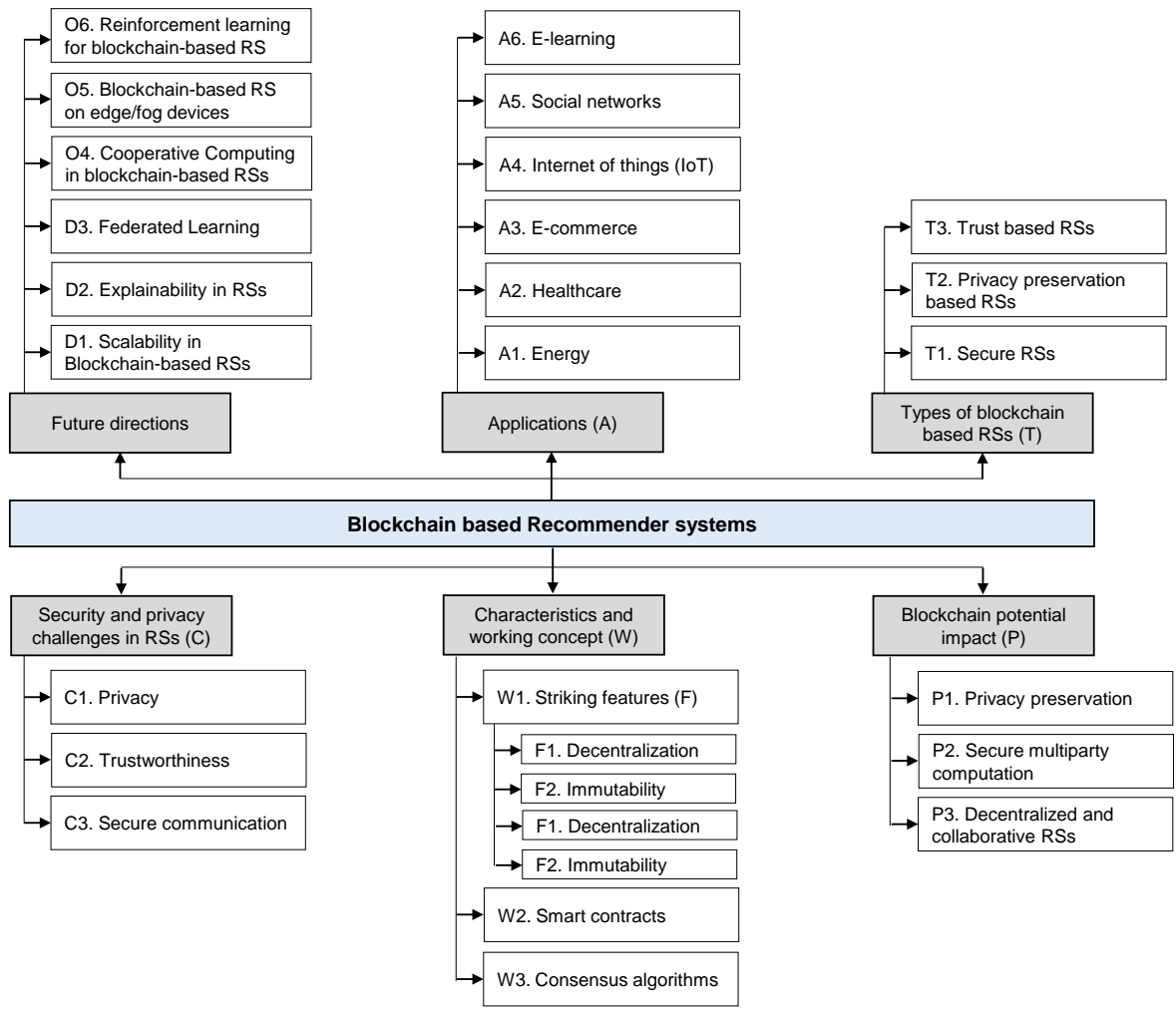

Figure 2: Taxonomy of blockchain based RSs.

A closely related concept is that of trust. In the context of RSs, trust has multiple meanings. First, it is related to trustworthy and secure services - i.e. the users trust that the RS handles their sensitive identifiable data properly and protects their privacy. Second, the users should trust the generated recommendations for their precision and correctness Wang et al. (2018), in order to increase the chance of recommendations to be accepted. Finally, there is a third definition of trust in the context of trust-aware recommendations Eirinaki et al. (2014). This refers to social RSs that leverage the circle of trust of the user to enhance the quality of recommendations. The focus of this work is on the security and privacy-related meanings of trust so the third meaning will not be elaborated further.

*C1. Privacy:* In RSs, preserving the privacy of users is of utmost importance, because the RS generally gathers and stores various sensitive data related to the identity, usage of utilities, preferences, social information, etc. Zhan et al. (2010). If a potential intruder gains access to the RS engine, it would be possible to mine and steal sensitive user data, such as preferences, habits, location, etc. This problem becomes even more prevalent in applications such as healthcare, where the data stored might include even more sensitive information Kaur et al. (2018); Xu et al. (2019).

Authentication is considered a key factor in almost any knowledge-based system, and is the first layer of security in the digital world. It also has an important contribution in creating a trustful relation between the users and the RS. To access the services of a RS, a user should first connect to the RS network and authenticate themselves to the RS platform, which inspects the reliability of claimed identities by matching different user-related information. However, another layer of vulnerability is introduced as most RSs and the respective data are deployed and stored on third-party cloud frameworks. This may increase the risk of users' privacy violation.

The subject of privacy protection and security in RS has been studied from various aspects including the legal and regulatory framework, and the technology-related data encryption and distributed storage and processing techniques Xu et al. (2018).

**C2. Trustworthiness:** The precision and correctness of the recommendations can deteriorate due to several reasons, and mainly due to the training process itself. Even in cases where everything is done properly, the quality and trustworthiness of the recommendations can be compromised, either due to bias existing in the data, or due to malicious attacks. Despite the many and clearly defined algorithms employed by the recommendation engines, users are generally not provided with enough data and convenient reasoning that will allow to comprehend precisely why and how an item/action is being recommended to them. This leads to "black-box" models, which may hide inherent biases in the training dataset or the overall design that might go undetected. Explainability of recommendations is a prevalent challenge not only for RSs, but for all ML models in general Li et al. (2020b); Zhang and Chen (2020).

Finally, and most related to security and privacy, the dataset and/or the model might be compromised by malicious attacks aiming to achieve a particular outcome in the recommendation process. RSs can suffer from different kinds of malicious attacks launched by hackers who would try to leak the identities of the authenticated users and bias the recommendations, such as shilling attacks Cai and Zhang (2019); Zhang and Wang (2020), adversarial attacks Cao et al. (2020b); Deldjoo et al. (2020a), profile injection attacks Williams et al. (2007); Zhang and Zhou (2014), and poisoning attacks Fang et al. (2018); Hu et al. (2019).

Informally, the robustness of a RS measures its ability to prevent attackers from manipulating its output. When such attacks are targeted to rating-based RSs, they usually include injection of fake information on a user profile (e.g. fake ratings), generation of fake profiles that intentionally rate specific items higher than others (which are randomly assigned low ratings), or attacks to nuke specific items. To that end, robustness, which refers to delivering stable, secure and tailored recommendations, is becoming one of the most challenging security issue for RSs Hurley (2011); Mobasher et al. (2007).

A variety of strategies that can be employed to build RSs in a more robust way have been proposed. CAPTCHA has been successfully used in several online systems to identify humans vs. robots and can be applied in the RS context as well Ahn et al. (2003). Social trust Golbeck (2008) has also been widely explored Resnick and Sami (2007, 2008) as a way to identify adversary behaviors. A variety of robust matrix factorization methods for designing attack-resistant RSs have also been proposed Mehta and Hofmann (2008); Mehta and Nejdl (2008). An overview of attack-resistant RSs can be found in Aggarwal (2016) while a very recent survey of attack-detection approaches for RSs has appeared in Rezaimehr and Dadkhah (2021).

**C3. Secure Communications:** Although RSs need to implement the minimum security requirements, communications between users and the RS engine should implement all known security practices. RSs interact with users by sending and collecting different kinds of data. Moreover, data might be transmitted to remote cloud servers if the RS requires important computation and storage resources Alsubaei et al. (2019). Consequently, in RSs, the data communicated between a RS and users is usually assumed to be correct and authenticated. However, in practical scenarios, this assumption could be wrong since malicious users can manipulate the information of authenticated users or bias the recommendations for their own benefits O'Mahony et al. (2005).

## 2.2. Blockchain-based RS: characteristics and working concept (W)

RSs, initially appearing as content personalisation solutions for the Web, quickly expanded to the real-life with many applications ranging from health and wellness, to tourism and energy efficiency. The last few years, IoT have created a new landscape for applications (in smart cities and homes, healthcare, agriculture etc), where recommendations can be useful for many reasons. The ability of IoT to act as or on behalf of human beings further expanded the applications of IoT-based RSs and raised the need for human trust to the generated (and directly actionable) recommendations.

The popularity of RSs and the profit that they can bring to the sellers of the recommended products, has increased the interest of sellers to bias recommendations and manipulate the RSs for their benefit Adomavicius et al. (2019). Since trust is one of the main requirements for the acceptance of RSs technology, it is important to establish trust both to the providers of information (e.g. users, sensors, etc.), to ensure that they do not introduce bias Mohammadi et al. (2019), and to the recommendation algorithms, for being robust to adversarial attacks Du et al. (2018).

Blockchain offers a tamper-resistant ledger, where all transactions are recorded in a decentralized network to guarantee anonymity, integrity, and data security. Its characteristics may be used for RS development, where privacy and trust are key values for the decision making process, and can go beyond existing protocols for privacy-preserving data collection Beg et al. (2021). For example, in healthcare, users wish to share their sensitive information in exchange of better medical recommendations, but still ask for privacy and protection of their data from malicious adversaries (e.g. confidentiality, protection from forgery or modifications) Cao et al. (2019). Similarly, in social network RSs, users may require privacy of their context or preferences (e.g. they do not want to expose their exact location, or the exact items they viewed or liked), whilst they still want recommendations for similar items, actions or places Polatidis et al. (2017).

**W1. Striking features:** Blockchain networks are governed by the following properties:

- Continuous availability: Unlike traditional servers, blockchain practically never stops working, neither due to failure, nor due to maintenance.

- Reliability: The system completes its operations consistently and successfully, while providing explanations for possible transaction failures.

- Openness: The blockchain does not single out specific users or computers. The operating rules are open and transparent, since blockchain also relies on using trusted mathematical algorithms for regulating the behaviors of transactions. Moreover, exchanging information between nodes in the blockchain systems does not necessitate mutual trust. On the other hand, this feature depends on the type of blockchain, where three main scenarios are identified: (i) public blockchain, where it is completely open; (ii) consortium blockchain, where it is open to particular groups or organizations that have been given permission to access it; and (iii) private blockchain, which is only open to an entity or a specific user with a completely internal control.

- Security: At the level of each transaction, blockchain ensures that the property stays and is transferred to the right users. As for the operation of the entire system, blockchain protects users from theft, unauthorized access, double spending and fraudulent transactions. In doing so, blockchain relies on the concepts of cryptography, consensus, and decentralization, where data is organized in blocks and every block includes a transaction (or bundle of transactions). Every new block is linked to all its previous blocks using a cryptographic chain, where it is nearly impossible to hack it. Moreover, all transactions within the blocks should be approved and decided by a consensus mechanism, which ensures that every transaction is correct and true. Asymmetric encryption algorithms are mainly applied in the blockchain, where they are responsible on completing encryption/decryption bay using two asymmetric ciphers, the "public key" and the "private key". Commonly used asymmetric encryption algorithms in blockchain are Rivest-Shamir-Adleman (RSA) and elliptic curve cryptography (ECC) Chandel et al. (2019).

- Durability: Even in difficult conditions, the blockchain can confirm, but also properly transfer ownership of its data, while being resistant to a wide range of attacks.

- Final Stability: It can be seen from different perspectives. Indeed, due to the way the blockchain works, there are some cases where the answers given by the system are not constant, but over time the whole system eventually returns stable answers. Put differently, stability also refers to the case in which there is a positive rate of block arrivals and where blocks can be approved for a fixed peer-to-peer network (P2P). In this regard, the stability of blockchain-based RSs aims at enabling an external observer in determining, in finite time, which blocks will eventually be approved.

- Integrity: The behavior of the system does not include logical errors. The blockchain maintains the integrity of the data and ensures the security of transactions, as well as their history.

Blockchain technology offers many advantages compared to traditional database management systems (DBMSs) for storing data. The characteristics may vary between private consortium and public blockchains Mohan (2019); Zheng et al. (2017), but they can be beneficial for RSs.

***F1. Decentralization:*** The blockchain technology is characterized by a decentralized form of consensus between counterparties, with low transaction and control costs as opposed to traditional centralized systems. The blockchain contains information, similar to a database, but in a secure way that maintains data in a P2P network. In other words, the blockchain is a combination of computers connected to each other instead of a central server, which means that the whole network is decentralized. The peers that participate in the blockchain and keep record of the transactions are the keepers, who operate with specific protocols and perform mining tasks for allowing the chain to expand with new nodes.

In decentralized blockchain networks, the users who keep the transaction records vote on the historical events that they think are correct and according to these votes the users (agents) of the system decide the status of the blockchain using a consensus algorithm. The decentralized nature of public blockchains (e.g. Bitcoin, Ethereum) means that network participants must be able to reach an agreement on the common state of the blockchain. Unanimous consensus between network nodes results in a single blockchain containing verified data, such as network-confirmed transactions. However, network nodes cannot always reach a unanimous consensus on the future state of the blockchain. This leads to forks that are all valid. In the decentralized trading system, the creation of blockchain promoted a different way of recording and processing information that does not need to be trusted between the different entities involved in them while reducing costs, i.e. blockchain removes the need for trust between entities as they are subject to the authority of the technical system that they are convinced is permanent, indelible, and unalterable.

The recording of decentralized journals must not be possessed by deceptive acts of individuals participating in the network. To avoid this and prevent the counterparties from taking such actions, penalties are usually imposed on the wider social network. Fines range from loss of revenue to natural costs associated with the global network resources. In centralized payment networks, unanimity is achieved with confidence in the recording of transactions. This trust can be achieved either because extra money is given to secure genuine transactions or because penalties are imposed in the event of fraud.

Decentralisation applies well to the case of RSs when privacy Duriakova et al. (2019) or scalability Sardianos et al. (2017) are the main concerns. In contrast to data obfuscation, anonymisation, differential privacy, and encryption, which are usually applied to centralized systems, decentralized storage and processing of information used by the RSs (e.g. user ratings) eliminate the need of users to exchange the actual information with other users or a centralised node. Instead, they can share only the minimum information for the recommendation algorithm to operate. For example, in the case of matrix factorization (MF), the users share only the loss gradients of local MF models Duriakova et al. (2019). When it is not possible to guarantee unlimited processing and memory resources for storing and processing user ratings, distributed solutions can be employed to distribute the load among lower capacity peers, each one having access to a smaller network of users and the information they share Sardianos et al. (2017).

*F2. Immutability:* Immutability of records is one of the most useful characteristics of blockchain technology, since it protects the input data of the recommendation algorithm from unwanted changes Hofmann et al. (2017). Immutability facilitates the auditing process and brings more trust and integrity to the data stored in the blockchain.

In order to store a transaction in the blockchain network, a timestamp is added to it embedded into a "block" of information that also contains the hash of the previous block in the chain, and a hashing process cryptographically secures it. When a block is tampered, the blockchain breaks, and the reason for this break is directly available. For example, in the case of collaborative filtering algorithms the main input for the RS are the ratings of users for items (known as the utility matrix). An adversarial attack may target the utility matrix in order to modify ratings and affect the recommendations. Using a public blockchain the ratings will be stored on a large number of nodes, thus making it nearly impossible to tamper with them. In a similar manner, when content-based or knowledge-based RSs are employed to tackle the sparsity of ratings, there is still the option for an attacker to tamper with the user or item profiles by adding or removing content, and consequently bias the recommendations. The use of blockchain for storing profile information can take advantage of the immutability property and protect this information from unauthorised modifications.

*F3. Auditability:* Auditability is an official examination and verification of transaction records performed by third party experts either internally or externally. The ability of blockchain to act as a distributed ledger, where transactions are stored and validated by the participants on a continuous basis, has shifted the centralised audit paradigm towards a new self-auditing format Broby and Paul (2017).

The main problem in digital auditing is how to achieve the consent of all parties and verify the history of events/transactions. Centralized trading networks capitalise on trust to a centrally controlled transaction auditing mechanism. The blockchain trading system promoted a different way of transaction auditing that does not rely on trust between the involved entities. This trust can be built by investing to secure genuine transactions or by detecting and penalizing fraud. In a distributed ledger like blockchain, auditing brings new challenges Liu et al. (2019). The malleability of transactions leaves space for transaction modifications which are hard to trace Andrychowicz et al. (2015). Another challenge for auditors are blockchain forks, where two or more groups decide to recognise only their own blocks as valid (e.g. because all other blocks differ in size, or follow different rules). In that case, auditors must be aware of long or short-term forks in order to properly verify the transactions. A major challenge for auditors relies on the way transactions can be performed, using only the necessary private keys, from any point of the network. This makes it harder to notice all transactions that take place and validate their origins.

*F4. Fault tolerance:* Fault tolerance is the ability of a system to continue to operate even in the presence of errors. In general, fault tolerance is related to the concept of reliability, i.e. the continuous uninterrupted operation of a system, without fail. A fault tolerant system should be able to manage both hardware and software faults, power outages, and other types of problems that are unpredictable in order to continue to meet its specifications. Fault tolerance can be based on redundancy of processes and resources, failure masking, and recovery using self-inspection and self-healing capabilities Egwutuoha et al. (2013). Reliability is the ability of a system to avoid failures, which occur more frequently or are more severe, and also avoid longer downtime, which is not acceptable to its users. Otherwise the system is considered unreliable, i.e. it does not persuade its users to use it.

When a faulty service fails equally for all users and all of them face the same consequences, then the failure has homogeneity, e.g. an inactive server computer displays a similar message in browsers. When the service fails only to a few users, and yet there are some other users who continue to have the right service, then we have heterogeneity in failure, which is often referred to as the Byzantine failure and is one of the most difficult problems to solve in modern fault tolerance systems. The difficulty mainly lies in the detection of the fault, since the service status is constantly changing.

One of the main features of blockchain technology is its ability to form consensus of all participants upon a decentralized and shared transaction log. This decentralisation of trust is achieved by combining cryptography with smart incentives that allow participants to agree on the state changes of the ledger. The distributed nature of blockchain that allows nodes to temporarily fail, and the consensus that is achieved to a common state of the ledger, even when nodes temporarily fail, makes it an ideal platform for developing Byzantine Fault Tolerant (BFT) solutions Lamport et al. (1982).

In private or restricted blockchain environments, where an individual enterprise or a credible authority is responsible for the blockchain fault tolerance, the fully BFT may be unnecessary. In situations like this, a crash fault-tolerant (CFT) consent protocol may be more than sufficient. In a multilateral, collaborative and decentralized use case, novel security methods that guarantee confidentiality, integrity, and availability (CIA) are needed. BFT protocols that establish traceability and accountability of the various actors are the required tools for achieving trust Stifter et al. (2019).

Trust is one of the factors that affects user acceptance of RSs Pu et al. (2012) and can build on the competence and usefulness of recommendations, as well as on the integrity and consistency of the RS. Fault tolerance can be beneficial in the latter direction, since it will guarantee the uninterrupted and benevolent operation of the recommendation engine.

**W2. Smart contracts:** The advent of Ethereum blockchain in 2015, introduced the concept of *Smart Contracts*. Smart contracts are computer programs that codify the business logic behind an agreement, which are activated and executed when certain conditions are met automatically. They can be securely deployed in blockchain, in order to guarantee to all sides of the agreement that when all the pre-set conditions are met, the contractual terms will be enforced automatically and will always produce the same result for all parts of the agreement. Smart contracts are automatically activated and the transactions are recorded in the blockchain making the information unchanged and unquestionable. This removes the need for a third party entity to act as an arbitrator, and avoids the influence or manipulation that this entity could have applied to the execution of the contract. These properties add to the reliability and availability of smart contracts and make them ideal tools for decentralized applications (DApps).

Smart contracts are an important element in processing information in the blockchain. If the blockchain represents a secure process of distributed information storage, then smart contracts complement this process by integrating it into a standalone Turing machine. When the programming language is Turing-Complete, which means that it allows the calculation of complex operations and functions, smart contracts can execute and process data on the blockchain executed either automatically or after an external stimulus that may come from an IoT device or from an external ERP program. For example, for a truck that transports sensitive products from point A to point B, the transport contract states that the temperature should not rise above -10 degrees Celsius. If for some reasons this temperature rises and becomes -8 degrees Celsius then automatically a sensor informs the blockchain and a smart contract "wakes up" and records this movement. The final recipient is aware that the temperature has risen above limits and can exercise penalties as stipulated in the agreed contract. The fact that the smart contract code runs on the blockchain makes it almost impossible to falsify information. Also, the distributed structure of the program and its simultaneous execution on all nodes of the network prevents its execution by a malicious or a compromised node that has been victim of a cyber-attack.

Smart contracts can be beneficial in the case of distributed machine learning frameworks, where participants share data and local models that require resources for being trained. Using smart contracts, participants can share their models publicly on the blockchain, where they will be free to use for inference. Primarily, this allows to create a value chain of shared resources and secondly it leaves space for profit, to the model providers, using rewarding schemes Harris and Waggoner (2019). Smart contracts can be used to check whether data or model updates (in the case of incremental algorithms) have improved the model performance that is periodically evaluated against a test dataset. Based on the contribution, each partner is rewarded at the end of this process. A distributed matrix factorisation scenario, where various partners are sharing their local models can be an ideal application of smart contracts in RSs Lisi et al. (2019).

**W3. Consensus algorithms:** Proof of Work (PoW) is the oldest of consensus algorithms. It was first implemented in Bitcoin, but the real idea has been around for quite some time. In PoW, validators (referred to as miners) record the data they want to add until they produce a specific solution. The PoW protocol sets the conditions for what makes a block valid. For example, only a block whose hash starts with 00 will be valid. Miners can modify a parameter in their data to produce a different result for each guess, until they get the correct hash. In large blockchains, the stake is quite high. To compete with other miners, one needs a warehouse full of hardware that would deliver high computing power to build a valid blockchain. The reward, in mining, is the cost of these machines and the electricity required to operate them. The only way to recoup the initial investment is to do mining, which is a great reward if a new block is successfully added to the blockchain.

Proof of Stake (PoS) was proposed in the early days of Bitcoin as an alternative of PoW. In a PoS system, no miners, specialized material, or bulk application are required. PoS does not rely to external resources (electric power or harware), but on an internal mechanism - the encryption. The rules differ in each protocol, but there is generally a minimum amount of money that a user must possess to meet the entry criteria. More specifically, validators' money are locked in wallets and validators bet on the next block of transactions that will enter the blockchain. When a block is selected by the protocol, stakers receive a percentage of the transaction cost, depending on the amount they have bet (i.e. locked). The more money they have locked in, the more they earn. If they try to cheat by proposing invalid transactions, they will lose a part (or all) of their share. So PoS relies on a mechanism similar to PoW, where honesty is more profitable than a dishonest behavior.

Many alternatives to the above have also been proposed in the literature and have been adopted by other networks, such as the Delayed PoW, Leased Proof of Stake Consensus, Proof of Authority, Proof of Burn, Delegated Proof of Stake, Hybrid PoW/PoS Consensus Sankar et al. (2017). The main concept in all these consensus mechanisms is that stakeholders create new blocks and are rewarded for this task, whereas other nodes are verifying these blocks, which are added to the blockchain. In order to secure the network and the information it holds, the cost of changing information is increased since multiple nodes must be modified in order the change to happen.

### 2.3. Blockchain's potential impact on RSs (P)

The concept of trust is of utmost importance for RSs, especially when collaborative filtering algorithms are employed. As explained in the introduction, it is of equal importance for the RS to keep malicious users outside of the network, and to guarantee that the information used by the recommendation algorithm (e.g. ratings), the learned models, and the produced recommendations are protected from adversarial attacks. Blockchain and smart contracts can be beneficial in this direction Yeh and Kashef (2020). Security and trustworthiness are two features that blockchain offers using the PoW and the immutability of the ledger respectively.

*P1. Privacy preservation:* In RSs, recommendations are generated based on users' past data. To design accurate and efficient RSs, large volumes of high quality data are required. Consequently, companies and RS developers are gathering massive quantities of user data, which are usually very sensitive. However users' privacy concerns are often ignored. Companies and RS developers develop various risk reduction strategies for addressing the privacy problems, including cryptographic security, data anonymization and obfuscation. Nonetheless, user data are still subject to various security and privacy issues, such as transaction linkability, recovery of encryption keys, etc. Blockchain offers novel crypto-privacy methods to further strengthen data security and privacy, thus allowing users to become anonymous and protect their profiles from being exposed. Accordingly, to preserve the user privacy and increase the security, blockckain uses different cryptographic strategies, among them: (i) cryptographic hash functions, which help in creating a unique output (as digest) to every data input (of any size) Lu (2019); (ii) asymptotic cryptography approaches(olso known as public key cryptography) which deploy two keys, i.e. a public key and a private, key that are mathematically related to each other He et al. (2017); (iii) cryptographic addresses, which are alphanumeric sets of characters derived from blockchain network users' public keys based on cryptographic hash functions Li et al. (2018). Decentralisation of data is one of its advantages, which minimizes the risk of a user profile to be completely retrieved even if one or a few nodes are compromised. Keyless signature infrastructures (KSIs) and SSL certificate devices that have been developed for blockchain allow users to verify the integrity of their data and signatures without revealing their keys. All these, along with the resilience of the blockchain P2P architectures are guarantees for a secure and seamless data processing Niranjanamurthy et al. (2019).

Data tampering is among the most fascinating privacy protection problems. Customers need to exchange their personal details with these online company sites for different individual or business purposes Al Omar et al. (2019). Taking advantage of this exchange of knowledge about a person, online business sites collect client details, including the most confidential client information, to conduct a separate data collection without the permission of the client. To some degree there is a shortage of safe monitoring of client details in such applications. Blockchain technology guarantees confidentiality in the processing of data for clients on these web platforms since it is a secure public ledger for storing data transactions. With the use of smart contracts, blockchain-based applications can guarantee the use of data for specific purposes only (intent) and thus increase the trust of their users McKinney et al. (2017).

*P2. Secure multiparty computation:* The penetration of information and communication technology (ICTs) in our everyday lives led to a data explosion Casino and Patsakis (2019) and fed people with more choices that they can handle or probably like. In order to filter information, we need more personal data, which raises privacy issues for these schemes. Although autonomous advisors can protect privacy, they do not have the required efficiency to be broadly implemented. Secure Multi-Party Computation (MPC) provides improved anonymity and precision Zhou et al. (2021), so if it is properly combined with blockchain, it can overcome privacy and confidence challenges Bosri et al. (2021); Frey et al. (2016a).

*P3. Decentralized and collaborative RS:* As a decentralized infrastructure, blockchain is regarded as substitute to the shared centralized data storage systems Wang et al. (2019a). Authors in Harris and Waggoner (2019) developed a platform for collaborative dataset creation and continuous ML model update with the use of smart contracts. The model is then made openly available on the blockchain for inference. In order to ensure the consistency of the model with respect to any test collection, both financial and non-financial (gamified) compensation mechanisms can be promoted, for the provision of good results. The Ethereum blockchain[1], can be the basis for a similar implementation.

Authors in Arif et al. (2020) present a data-sharing framework scheme that employs a blockchain-based decentralized network, where each node can be linked directly to other nodes, in order to facilitate data exchange between nodes. Each form of node has a unique structure and path for the data communication. When the user node sends the destination evaluation data to the server node, the server node sends the data from the machine learning method to the user node. The sensor transmits dynamic data to the user node as a factor for finalizing the recommendation generation phase. In the process of transmitting results, each node in the blockchain network performs a number of functionalities, involving chaining block, hashing, and broadcasting.

In Wang et al. (2019a), the authors introduce a decentralized knowledge graph creation approach. They employ crowdsourcing methods, supported by blockchain-powered smart contracts in order to maintain openness, accountability and auditability. The resulting decentralized knowledge graph is the basis for a deep recommendation framework, and several case studies are used to confirm the efficacy of the system. The authors intend to provide a new, decentralized method to building the knowledge graph and serve as a guide and direction for potential study and realistic use of the knowledge graph. Fig. 3 portrays the flowchart of the deep RS based on the decentralized knowledge graph. The overall model can be split into two main components, i.e. the on-chain component and off-chain component. The former focuses at building a domain knowledge graph through crowdsourcing and that is activated by blockchain-powered smart contracts, while the latter employs the decentralized knowledge graph for producing the recommendations.

---

[1] https://github.com/ethereum

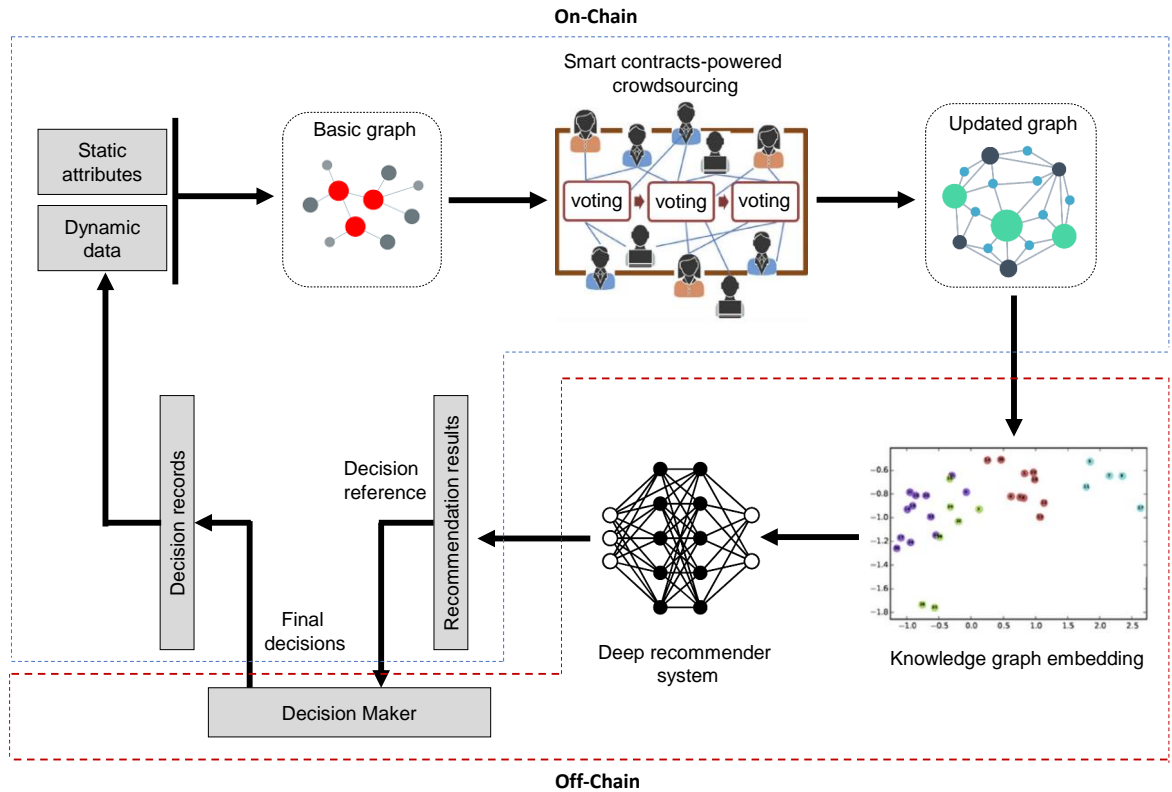

**Figure 3:** Flowchart of the deep RS based on the decentralized knowledge graph Wang et al. (2019a).

The authors of Arora et al. (2020) bring into consideration the facets of blockchain and deep learning elements of the market as the future of artificial intelligence in the business scenario. The report also includes the advantages and risks of utilizing artificial intelligence in industry. The study collects results from the usage and deployment of blockchain and deep learning components.

In Wibowo et al. (2019), the authors propose a method for automating the security of the Internet utilizing white-list packet filtering. In order to decide the filtering laws, smart contracts are used to provide protected communication media. Meanwhile, the controller streamlines packet filtering by implementing the defined forwarding rules in the network switches. The authors also evaluate white-list packet filtering using Mininet for network emulation and Ethereum Rinkeby networks for smart contract implementation. The outcome indicates that the suggested white-list scheme is capable of filtering the packet without resulting in excessive latency.

### 2.4. Blockchain-based RS architecture

The main architecture of the blockchain can be decomposed into three layers, namely the Protocol, the Extension, and the Application layer. The Protocol layer implements the P2P protocol, the consensus algorithm, and data storage. In particular, it consists of two components: i) the storage layer that stores all the user-item transaction data in the blockchain, and ii) the network layer that contains the peer nodes that run the distributed algorithm and carry the encrypted signature. The encrypted transaction blocks can be stored in private blockchain services such as Hyperledger Fabric, or take advantage of cloud-based solutions such as Microsoft's blockchain-as-a-service or Amazon's managed blockchain Lu (2018).

On top of the Protocol layer is the Extension layer that can implement the business logic and algorithms of the RS using smart contracts. Ethereum or Hyperledger Fabric can be used for this task and they should offer the required functionality in the form of APIs. One or more RS algorithms can be accessed, such as the memory-based, user-based Resnick et al. (1994) and item-based collaborative filtering (CF) algorithms Sarwar et al. (2001), the model-based matrix factorization models such as the original SVD-based one Koren et al. (2009) and its variations (e.g. SVD++ Koren (2010) and non-negative matrix factorization Gillis (2020)), as well as models are much more flexible in terms of adding additional sources of information on top of ratings, such as social recommender algorithms Gulati and Eirinaki (2019); Jamali and Ester (2010); Ma et al. (2008) or the embeddings-based factorization machines Kula (2015).

The definition of the information that is used to model users, recommendation items/actions and prior knowledge, and the representation model employed for storing user preferences, behaviour and feedback (e.g. a utility matrix, a social graph, a knowledge graph, a rule database, a deep neural network, etc.) are two main decisions that affect the implementation of the RS. In addition, the decision on whether, when, and how the model is updated sets requirements that affect the architectural choices. The last and more important decision is on which recommendation algorithm to use. The algorithm properties, e.g. if it is incremental, online, distributed, synchronous, etc., as well as specific characteristics with respect to what types of data input it accepts, and other concerns, such as ratings sparsity etc., are of big importance in making such a decision.

Apart from the functional requirements of an RS, which mainly focus on the quality of the generated recommendations, there is a long list of non-functional requirements that are important for the successful delivery of recommendations. They include adaptability, response time, scalability, performance, security and privacy, among others. The majority of privacy solutions for RSs mainly rely on centralized schemes, which raises privacy concerns. The decentralization of user-related data can be beneficial for scalability Sardianos et al. (2017) and user privacy Berkovsky et al. (2007). However, it can affect the quality of recommendations, which is strongly related to data sparsity and representativeness of users and items in each partition.

The protection of user privacy can be achieved by: i) defining an access policy to data and models, ii) maintaining the statistical properties of data whilst anonymising the original data in a way to avoid their revelation, and iii) using cryptography techniques to mask data without affecting the quality of the algorithm results.

Blockchain is a decentralized ledger of all transactions across a P2P network. By adopting the blockchain technology in RSs, users are able to approve recommendations or transactions without the necessity of a central certifying authority. Fig. 4 illustrates a general flowchart of a blockchain-based RS. Typically, one of the main assets of recommendation systems is the data contributed by users or item providers in the form of profile, descriptions, feedback etc. Therefore, users can use blockchain for storing sensitive information needed by RSs, e.g. preferences and selected items, digital transactions, special needs, personal identifying information, rating feedback, etc., which will be encrypted and secured. The company/RS owner gathers and stores other useful data for the RS in the blockchain network for avoiding hackers' attacks and any other dangers. In matrix factorisation, which is the state-of-the-art approach in collaborative filtering algorithms for RSs, the utility matrix $R$ of ratings provided by a set of users $U$ for a set of items $T$ is the key knowledge employed by the algorithm. The matrix $R$ is factorised in order to obtain user and item profiles and make recommendations based on the approximations of missing ratings. With the use of blockchain the user-provided ratings are encrypted before being sent to the recommendation engine and the same holds for the matrix derivatives, with the use of a crypto service provider (CSP). As depicted in Algorithm 1, the process begins with collecting (lines 1-5) and encrypting each rating of a user for an item using the public key of the CSP. The encrypted ratings are sent to the recommendation engine which generates random masks for each rating, encrypts them and sends them back to the CSP. In the matrix factorisation step (lines 6-12), the CSP decrypts the masked ratings (i.e. $r_{ij} + m_{ij}$), computes their d-dimensional vector representations, encrypts them and sends them back to the RS. The RS unencrypts the d-dimensional vectors, and removes the respective mask vectors to get the final approximations. The rating approximation process is repeated until a stopping criterion is met.

Blockchain technologies combine the advantages of their distributed nature with the use of cryptography in order to guarantee privacy and immutability of data. At the same time, using smart contracts blockchain technologies can guarantee the proper execution of the algorithms on the right data, thus adding to the security of the whole process. What is important in the case of any application that is ported on the blockchain, and also in RSs, is to properly define what type of information is stored in the blockchain, how it is encoded, and how this information is fed to the recommendation algorithm.

The amount of information exchanged between nodes, the amount and format of information stored in each node and the computation load of each node, affect the overall performance and privacy as well as the scalability of the solution. Moreover, choosing a blockchain platform is the first step in developing the blockchain-based RS, and for doing this is important to review the main characteristics of platforms, such as the block size, the block count in the main chain (chain height) and the speed of transaction processing. Bitcoin offers a block size of 1 million transactions, whereas it comes third in the transaction processing speed after Ethereum and Bitcoin.

### 2.5. Types of Blockchain-based RSs (T)

*T1. Secure RSs*

Secure RSs are generally built using a blockchain protocol supporting smart contracts and resolve different information management problems, e.g. integrity, security and authenticity, using the concepts of cryptography and hash functions Idrees et al. (2021). In this context, the authors in Lisi et al. (2019) introduce a smart contract-based RS (see Fig. 5) that relies on Distributed Ledger Technology Wood et al. (2014) and Ethereum TestNet. The principal main operations provided by the RS include registering new users (Register), creating new items/actions (Create item/action),

**Algorithm 1** A blockchain-based implementation of collaborative filtering algorithms that use the ratings matrix

1: **procedure** COLLECTUSERRATING($u_i, t_j, r_{ij}$) ▷ user $u_i$ rates item $t_j$
2:   User: encrypt rating $r_{ij} \to Enc.r_{ij}/$
3:   send $.u_i, t_j, Enc.r_{ij}//$ to the RS
4:   RS: mask rating $Enc.r_{ij}/ \to ER.r_{ij}/ = Enc.r_{ij} + m_{ij}/$
5:   send the masked and encrypted rating $ER.r_{ij}/$ to the CSP
6: **procedure** MATRIXFACTORISATION($ER.r_{ij}/$) ▷ Ratings' matrix $M$ is approximated
7:   CSP: decrypt $ER.r_{ij}/ \to .u_i, t_j, r_{ij} + m_{ij}/$
8:   set the d-dimensional vectors and compute their encryption $EV.r_{ij} + m_{i,j}/$
9:   send the encrypted $EV.r_{ij}, m_{i,j}/$ to the RS
10:  RS: decrypt $EV.r_{ij}, m_{i,j}/$ and remove masks
11:  update rating matrix approximation $R$
12:  verify stopping criteria

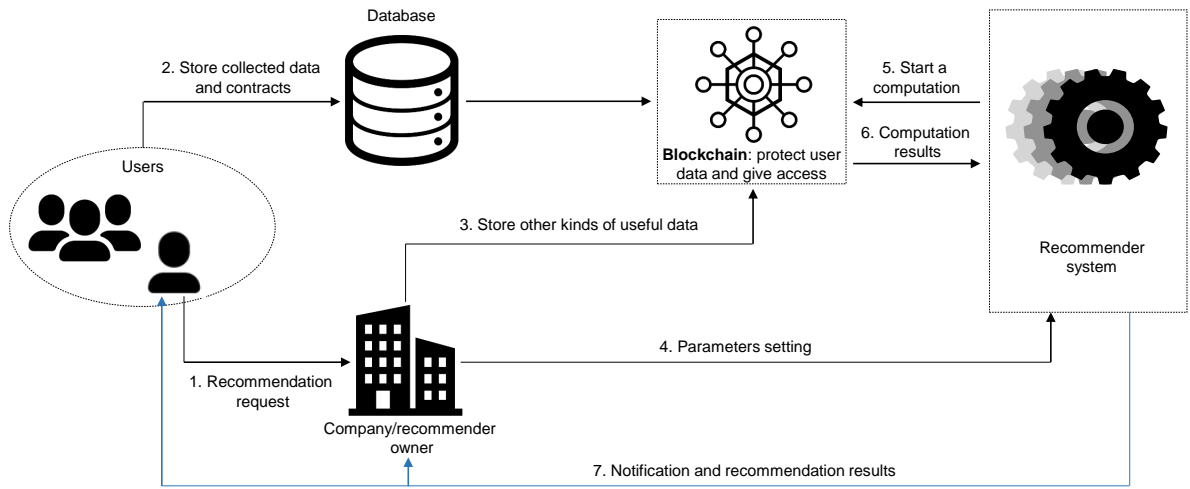

**Figure 4**: A general block-diagram of a blockchain-based RS.

rating existing items/actions (Rate item/action), and computing the scores of existing items/actions (Compute score). The RS allows decentralized ratings, ranking of several items/actions, and it is designed over a platform running smart contracts on a public blockchain, without any centralized authority. The capabilities of the framework have been widened via the integration of an authorization process, which gathers preferences exclusively from users eligible for rating the items/actions. To demonstrate the feasibility of the solution, a prototype of a decentralized recommendation system was designed and deployed on the Ethereum TestNet, and a series of experiments were included to determine its performance.

In a similar manner, in Khatoon (2020), a smart contract-based RS that enables managing medical data management and streamlining complex medical procedures, is proposed. In addition, a comprehensive study has been conducted to investigate the use of blockchain in healthcare workflows and the feasibility of smart contract-based frameworks in different use scenarios when realistic clinical databases are considered.

*T2. Privacy-preserving RSs*

Recommendations are generated by analyzing users' data. Consequently, a massive amount of high quality data are required to infer personalized and accurate recommendations. However, significant quantities of personal users' data are gathered, which are often highly sensitive and ignoring users' privacy concerns. Although in the past the developers of RSs have employed various privacy-preserving approaches to overcome these concerns, WAN (2018); Casino et al. (2019b); Zhan et al. (2010), none of them is able to guarantee cryptographic security. To bridge the chasm, the characteristics of blockchain in terms of supporting secure multiparty computation have been used in Frey et al. (2016a). In that way, potential users can allow RSs to generate personalized recommendations for them, disclosing their personal data. This results in reducing frauds and misuses, and a higher willingness for data sharing from the end users' point of view. Similarly, the authors in Bosri et al. (2021) propose a privacy-preserving platform for a recommendation system through the integration of artificial intelligence and blockchain. The platform gives the user

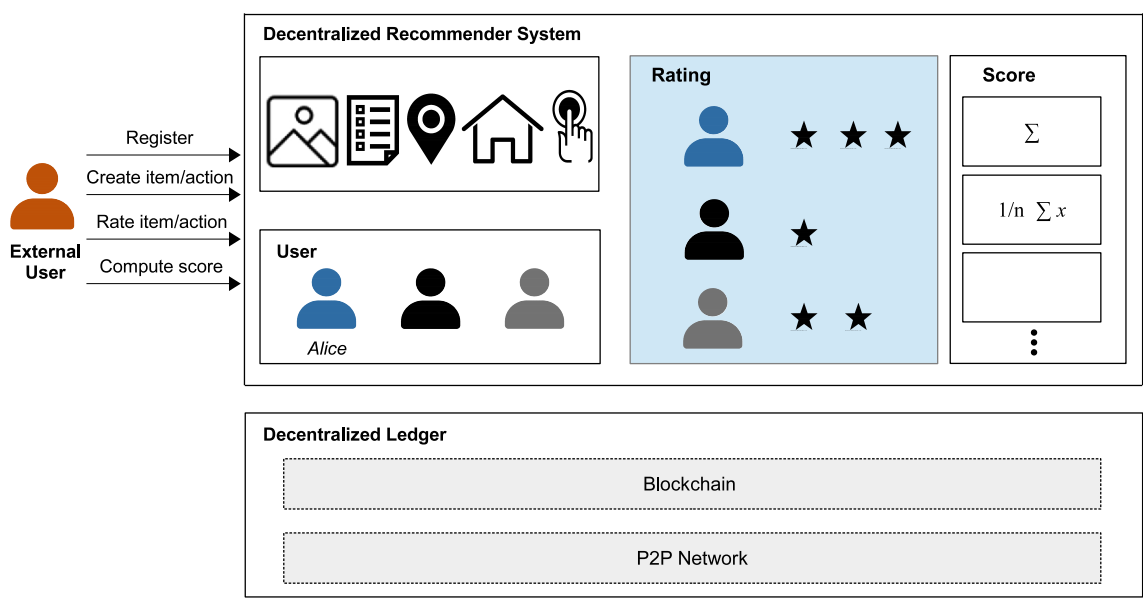

**Figure 5:** A decentralized blockchain -based RS proposed in Lisi et al. (2019).

a secure environment in which data can be used with the required permission.

Although decentralized RSs have good privacy protection capabilities, they still miss the required efficiency to be broadly accepted. Therefore, the authors in Casino and Patsakis (2019) have adopted blockchain as the core of a decentralized RS, where they have equipped it with a wide range of characteristics and in parallel, preserved users' privacy. The employed decentralized locality sensitive hashing classification, in addition to an ensemble of recommendation engines with reference to the manner in which data has been handled by users. Fig. 6 illustrates a typical flowchart of a secure RS based on the combination of blockchain and deep learning Arora et al. (2020).

*T3. Trust-based RSs*

Trust is an essential parameter to develop efficient RSs with accurate decision making. By using blockchain in RSs, it becomes possible to guarantee trust between users Mohammadi et al. (2019); Yeh and Kashef (2020). In this category of RSs, secure and trust-based frameworks are developed by the adoption of blockchain-supported secure multiparty computation and the addition of smart contracts within the principal blockchain protocol. Therefore, this helps in designing reliable RSs for collaboratively creating trust-based databases and hosting steadily updated models using smart contract systems. Moving forward, the accuracy of these models can be assessed using incentive mechanisms offering a completely trust-based RS with acceptable performance Porkodi and Kesavaraja (2020).

## 2.6. Applications of Blockchain-based RSs (A1)

*A1. Energy*

RSs in the energy sector aim to conserve natural resources, maximize users' savings, and to build a relationship of trust in which users share their data and the system influences their behavior. This interaction allows the system to collect more data of high quality regarding a user's real preferences, but also positively influences the users' future behavior toward more sustainable actions Alsalemi et al. (2021); Dahihande et al. (2020); Sardianos et al. (2020b). However, when designing energy efficiency based RSs, evaluating the recommendations' engine accuracy is a continuous process due to the instability of energy usage users' preferences. To achieve this, it is important to motivate users to contribute their activity data and adopt a data collection methodology that will feed the RS. Platform analytics, ratings, and user surveys are all good inputs for RSs but at the same time, they act as the indicators of recommendations' accuracy Sardianos et al. (2020a). An important goal is to guide users so as to gather their energy consumption feedback intensively through ratings and likes (explicit) but more importantly through usage and activity data, which is the asset for the $.EM/^3$ RS platform [2], which has been developed to encourage energy efficiency in buildings using behavioral change based on RSs, micro-moment analysis, and AI. Hence, this increases the data repository and the quality of extracted knowledge.

On the other hand, smart grids are expected to offer a range of advantages to both consumers and energy providers through smart metering and data sharing. That being said, smart meters creates a range of privacy problems if consumption data is analyzed. When sharing data, apart from privacy, it is important to ensure data reliability Rubio et al. (2017). Therefore, the development of secure blockchain-enabled smart meters is receiving an increasing interest as the actual utility meters enable easy manipulations to give incorrect readings and generate corrupt bills, which leads tremendous financial losses for utility companies. For example, in Bokhari et al. (2019), Bokhari et al. utilize smart meters in tandem with an Ethereum-based blockchain for increasin the security of energy usage data. While in Hassan et al. (2019), a blockchain based solution was designed for auctioning the micro-grid resources. To maintain the privacy and security of the auction, differential privacy strategies to ensure the disability of adversaries to access or alter individuals' private information.

---

[2] http://em3.qu.edu.qa/

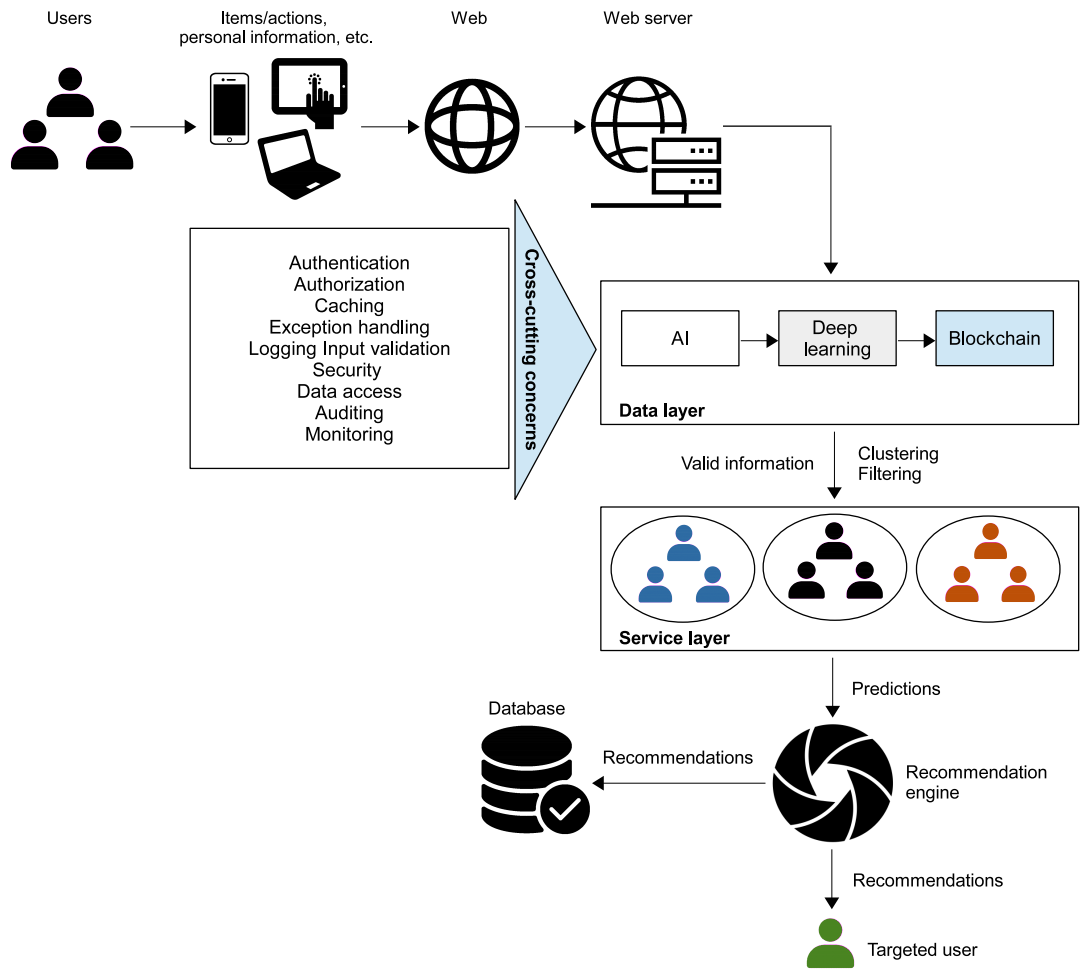

**Figure 6:** The overall architecture of a secure RS based on blockchain and AI Lisi et al. (2019).

Moreover, the public and legislators have recently voiced concerns about the security, privacy, and transparency of IoT sensing devices. Blockchain and distributed ledger systems have provided alternatives by transparently bridging two or more untrustworthy parties. In this regard, a new architecture, which incorporates a control system and automations in an intelligent building energy RS, was developed in Rahman et al. (2020).

*A2. Healthcare*

The vast amount of medical data generated every second have given rise to both a huge opportunity in analysing them in order to help healthcare experts to make patient-oriented decisions, and also to concerns related to privacy and security due to the high sensitivity of medical records. Therefore, the necessity of using RSs for helping healthcare experts make effective and precise health-related decisions has emerged Tran et al. (2020); Yong et al. (2020). In this context, the use of blockchain can help patients become true owners of their medical data and history. To that end, an increasing attention is put towards developing blockchain-based RSs for the healthcare domain. For example, in Bhardwaj and Datta (2020) the authors present an RS for patients with diabetes, which stores patient records in the blockchain, making them the true owners of their medical history. Blockchain allows P2P information exchange without the need of a reliable third party to verify transactions. The medical personnel gets access to the data, whereas patients can decide whether to share their data with more users (e.g. family, or caregivers). The RS employs information about a person's condition (e.g. heart rate, weight, sugar level, etc.) and lifestyle choices, and creates a knowledge graph. The graph is used to recommend activities (e.g. exercising, eating healthy) that promote healthier habits.

In Abbas et al. (2020), blockchain is used for securing all the data (i.e. data about suppliers, manufacturers, distributors, etc. and system assets and transactions) related to the drug supply chain management. Each transaction is validated by other users, appended to the blockchain, and executed. Using smart-contracts, the proposed architecture implements business terms and conditions. In order to avoid delays, the smart-contracts are deployed only to those network nodes that are responsible for the validation of transactions. The recommendation engine builds on top of patient reviews and comments about medicines. Using the positive/negative reviews and the patient medical condition, the engine is able to recommend the most appropriate and top-rated drugs for the specific patient conditions, using a content-based recommendation approach.

*A3. E-Commerce*

In most e-commerce-based RSs, recommendations are inferred from user data using collaborative filtering techniques. As a result, organizations are storing vast amounts of confidential consumer data. Given the sensitivity of such data, the consumers' privacy concerns should no longer be ignored Ali et al. (2020). Dealing with customer profiles while respecting people's right to privacy remains a major concern in e-commerce. Secure management of confidential data is vital to encouraging consumers to share their consumption data and enable businesses to make personalized recommendations Liu and Li (2020); Schuetz and Venkatesh (2020).

The work done in Frey et al. (2016a) presents a novel recommendation method leveraging the benefits of safe multi-party computing supported by blockchain. By applying the aforementioned method, any potential customer can allow an organization to use a recommendation algorithm without sharing their personal details.

Cognitive systems, IoT, and blockchain technologies are three areas that have resulted in various software development revolutions. It appears that combining these fields could contribute to the emergence of an intriguing field. Thus, a RS is developed to investigate the applicability of the suggested framework for IoT based on cognitive frameworks and block-chain technologies Wang et al. (2019a). Observed results verify the superiority of the developed RS over other options.

In Patel et al. (2020), KiRTi, a deep learning-based credit RS which uses blockchain for facilitating smart lending operations between prospective borrowers and prospective lenders is proposed. This helps in eliminating the requirement of third-party credit rating agencies for credit score generation. Authors in Lisi et al. (2019) suggest an alternative solution for developing a general-purpose RS that is built on Distributed Ledger Technology (DLT) to provide users with a more transparent ranking strategy. The proposed system graduates decentralized ratings, the ranking of various items, and is designed on top of a system that manages smart contracts on a shared blockchain with the absence of any centralized authority. The system's capabilities are also enhanced with the addition of an authorization module that extracts inclinations only from users who are authorized to rank the items. To demonstrate the feasibility of the solution, a prototype of a decentralized recommendation system was designed and deployed on the Ethereum TestNet, and a series of experiments were included to determine its performance.

The research presented in Frey et al. (2016b) suggests a novel approach for a future shopping system. A blockchain-based storage network protects customers' personal information. The system, which is based on the bitcoin protocol, can transact encrypted data in a tamper-proof manner and run secure multiparty computations with the input data only accessible by the data owner. As a result, a potential customer can allow a company to use functions such as a recommendation algorithm without revealing personal information. An entirely new shopping experience emerges when a low-energy transmitter (beacon) is used in conjunction. Based on encrypted personal data, the beacon automatically initiates a recommendation process.

*A4. Internet of things (IoT)*

Recent improvements in IoT have made it possible to implement smart city solutions (e.g. smart parking, traffic management, etc.) that employ RSs to provide suggestions to the users. The non-confidential sharing of user information with an untrusted/semi-trusted RS may infringe privacy, as user behavior and mobility patterns could be deduced by performing analysis of their prior actions. The authors in Saleem et al. (2020) present two solutions that retain the confidentiality of users patterns in parking RSs by employing k-anonymity (anonymisation) and differential privacy (perturbation) strategies whilst observing previous parking history. In particular, the k-anonymity method builds an anonymised database from an initial database containing users' preferences, while differential encryption disturbs the query response using the Laplace mechanism, rendering users indistinguishable in both methods and thus retaining privacy.

Blockchain, IoT, and cognitive technologies are some of the biggest technological advances in this era, which can be combined to provide smarter and more secure solutions. In this regard, in Porkodi and Kesavaraja (2020), existing variations of the aforementioned technologies are investigated, and a system for blockchain-based cognitive IoT is proposed for building trust over a reliable and secure RS. Authors in Putra et al. (2020) proposed the decentralization of intrusion detection systems (IDS) for networks, which uses blockchain to substitute the centralized processing with safe cooperation and solves some of the centralized IDS challenges.

*A5. Social networks*

Recently, huge quantities of information, such as text (tweets, comments, .etc), images, videos, and documents are shared on social networks Katarya et al. (2016). Due to the significant sharing of information through social networks, the information overload issue has been raised. Consequently various social network platforms employ RSs for overcoming this problem and suggesting valuable information to specified users Anandhan et al. (2018).

Using recommendation engines, different mobile applications can be suggested to mobile users. The vast volume of data and user information that is collected raises an issue of security and user privacy. In this context, a study of recommendations and trends in e-commerce technology and mobile applications is carried out in Umekwudo and Shim (2020). It explains how the blockchain can be integrated into the collective filtering recommendation systems so that users can configure their data protection and confidentiality.

In Jiang and Zhang (2019), Jiang et al. utilize the blockchain to develop an online social network (OSN)-based RS framework integrating the benefits of conventional centralized OSNs and distributed OSNs. Typically, by using smart contracts, the blockchain is considered as a trusted server for providing essential control services.

*A6. E-learning*

Nowadays, each aspect of the society is advancing and individuals are always looking for improvements for staying competitive in their careers. One of the significant technologies to achieve this goal is via adopting e-learning, which is significantly transforming the way in which learning is offered for learners De Medio et al. (2020); Maatuk et al. (2021). Navigating this wealth of information (which courses to take, from which instructors, institutions etc) has become a concern. In this context, RSs are used to assist learners with on time, tailored and engaging advice after an automatic identification of their preferences Tarus et al. (2018). In addition to courses delivered completely online, more and more in-person courses are accompanied by an online component, usually delivered via a learning management system (LMS). In this context, RSs can be used to generate recommendations on which courses to take next, how to improve one's grade, etc. However, such systems also store and process sensitive information (student IDs, grades, etc.). Existing technologies deployed in the e-learning platforms have raised some privacy and security issues (e.g. the leakage of private information, cybersecurity attacks, etc.), and hence resolving these problems has become essential for the users. In this context, the blockchain technology has the power to transform education and the use of RSs in e-learning environments in several ways Lam and Dongol (2020); Ullah et al. (2021). In Mikroyannidis et al. (2018) authors investigate the uses of Smart blockchain Badges (SMB) in data science education. They examine how SMBs will help learners with the intention of advancing their careers in data science by providing personalised tips based on their learning accomplishments. The study is carried out to improve data science accreditation by implementing a secure framework based on blockchain technology.

Table 2 summarises the various blockchain-based RS frameworks described previously and their characteristics, including the year of publication, classification, adopted methodology, targeted application and the main advantages of each framework.

## 3. Critical analysis and open issues

### 3.1. Comparative study

Although the increasing attention put towards the development of blockchain-based RSs, comparing the empirical performance of existing frameworks to inform the state-of-the-art is still very challenging and it is also difficult to faithfully comprehend the potential pros and cons associated with competing blockchain-based RS models. This is mainly due to various issues and aspects that are impeding the reproducibility of the empirical results and hence prevent a fair comparison of the performance under the same context, which they can be summarized as follows: (i) it is difficult (even impossible) to assess the generality of blockchain-based RS frameworks as most of them are evaluated on completely different environments, for performing different tasks and using different kinds of datasets (with different number of users, etc). Moreover, data is often preprocessed, resampled and/or augmented based on the specific need of each framework; (ii) there is no baseline available for comparison and most of the existing frameworks are just presenting case studies for illustrating the validity of their systems; (iii) there is a significant lack of comparative studies assessing the performance of blockchain-based RSs as this is a new topic and further effort is needed to create appropriate toolkits for comparison and benchmarking purposes. Accordingly, most of the existing studies were just evaluating their algorithms in comparison with their variants and/or versus trivial baseline approaches; (iv) distinct benchmarking metrics were employed in existing frameworks based on the targeted tasks and applications, thus making it virtually impossible to fairly determine the state-of-the-art.

In the reviewed works, datasets are used to evaluate and benchmark the developed RSs. In some contributions, whilst not published, raw data is collected and prepared by the authors and are used to evaluate recommendation performance. On the other hand, a number of proposed solutions utilize existing databases, such as the MovieLens dataset Casino and Patsakis (2019), the MNIST dataset Harris and Waggoner (2019), microblogPCU dataset Jiang and Zhang (2019), human resources job offer dataset Mikroyannidis et al. (2018), and e-commerce shopping data Frey et al. (2016a) among others. Such datasets are instrumental tools to understand the applicability of the proposed systems in real-life scenarios.

Despite the aforementioned issues, we attempt to conduct a comparative analysis of different blockchain-based RS models using what we have at hand as results. Table 3 reports the characteristics and performance results of the frameworks used in the comparison with regard to the publication year, implemented approach, computational time and recommendation accuracy. First of all, it seems obvious that all the approaches described in the table are proposed recently as part of the recent movement of the RS community to integrate blockchain into RSs. Thus, regarding the time computation, it can be clearly seen that it varies significantly from a framework to another. This is due to the type of the blockchain platform deployed, cryptographic algorithms used, targeted application, nature of RS developed, and also number of users. In this regard, Private-Rec [87] has achieved the lowest computation time since it adopts a simple collaborative filtering based RS along with a simple fully homomorphic encryption algorithm. In terms of the recommendation accuracy, the performance of all the frameworks exceeded 80% although there is a significant difference between the best accuracy achieved by KiRTi [123], which was 97.5% and the lowest accuracy reached by Abbas et al. [119], which attained 80.5%.

**Table 2**
Summary of the relevant blockchain-based RSs discussed in this paper and their characteristics with referencing their classifications and target applications.

| Work | Year | Class of RS | Method | Application | Advantages |
|---|---|---|---|---|---|
| Frey et al. (2016a) | 2016 | Secure RS | Blockchain-supported secure multiparty computation. | E-Commerce | Address privacy concerns and guarantee cryptographic secureness. |
| Mikroyannidis et al. (2018) | 2018 | Secure RS | Smart blockchain badges using smart contracts. | E-learning | Address the issues of using blockchain for data science education in industry. |
| Al Omar et al. (2019) | 2019 | Privacy-preserving RS | Privacy-securing platform using distributed ledger attribute of blockchain | E-Commerce (online shopping) | Ensure client information protection utilizing blockchain technology. |
| Lisi et al. (2019) | 2019 | Secure RS | Decentralized distributed Ledger technology platform | E-Commerce | Provides users with transparent and decentralized rating strategy. |
| Casino and Patsakis (2019) | 2019 | Privacy-preserving RS | Blockchain+decentralized locality sensitive | E-Commerce (movies recommendation) | Use blockchain as the backbone of a decentralized RS to ensure users' privacy. |
| Wang et al. (2019a) | 2019 | Privacy-preserving RS | Decentralized construction of knowledge graphs for deep RSs (blockchain-powered smart contracts) | E-Commerce (enterprise information) | Guarantee the transparency, integrity, and auditability using blockchain-powered smart contracts |
| Harris and Waggoner (2019) | 2019 | Privacy-preserving RS | Decentralized and collaborative AI on a blockchain platform | E-Commerce | Collaborative building of datasets and use of smart contracts to host a continuously updated model. |
| Jiang and Zhang (2019) | 2019 | Secure RS | Blockchain-based decentralized online RSs | Social networks | Provide safe, and privacy-aware functionalities of authentication and newsfeed notification. |
| Rahman et al. (2020) | 2020 | Secure RS | Combination of IoT, Software-Defined Networking (SDN) and blockchain | Smart energy management | Improve privacy accessibility, integrity, and overall security. |
| Porkodi and Kesavaraja (2020) | 2020 | Trust-based RS | Rs based IoT blockchain network with cognitive framework | Social networks | Increase the trust of the users in the RS and and guarantee a high-level of security. |
| Bhardwaj and Datta (2020) | 2020 | Privacy-preserving RS | Blockchain based platform with ML algorithms and optimization process. | IoT networks | Preserve the privacy of users and allow a secure sharing of data with healthcare experts. |
| Abbas et al. (2020) | 2020 | Secure RS | Blockchain and ML-based medicine supply chain management and RS. | IoT networks | Provide trusted and secure recommendations. and prevent recommending counterfeit medicines. |
| Patel et al. (2020) | 2020 | Secure RS | Combination of deep learning and blockchain-powered smart contracts. | E-Commerce (financial transactions) | Facilitate a trust based recommender ecosystem. |
| Yeh and Kashef (2020) | 2020 | Trust-based RS | Design secure databases using blockchain-powered smart contracts. | Social networks | Ensure High-level of security and and host steadily updated models. |
| Arif et al. (2020) | 2020 | Secure RS | Blockchain-based data-sharing system. | E-Commerce (online tourism) | Secure data circulation between nodes in Rss and handle data circulation from various nodes. |
| Lisi et al. (2021) | 2021 | Privacy-preserving RS | Decentralized rating framework based on Blockchain based Ethereum platform. | E-Commerce | Provide decentralization, review immutability and persistency. |

**Table 3**
Performance comparison of existing blockchain-based RS frameworks.

| work | Year | Approach | Time compelxity (ms) | Accuracy (in %) |
|---|---|---|---|---|
| Casino et al. Casino and Patsakis (2019) | 2019 | Blockchain-based privacy-preserving RS using a collaborative filtering architecture | - | 90.6 |
| Lisi et al. Lisi et al. (2019) | 2019 | A smart contract based RS built on top of distributed ledger platform | 2000 | |
| Private-Rec Bosri et al. (2021) | 2019 | AI-based privacy-preserving recommender system ensuring user data privacy. | 1.57 | - |
| Wang et al. Wang et al. (2019a) | 2019 | Decentralized knowledge graphs for deep RSs using blockchain-powered smart contracts | - | 83.3 |
| Arif et al. Arif et al. (2020) | 2020 | Blockchain-based RS that enables decentralized data sharing of tourism destinations | 203.6 | - |
| Abbas et al. Abbas et al. (2020) | 2020 | Blockchain and machine learning-Based RS for Smart Pharmaceutical Industry | 53 | 80.5 |
| KiRTi Patel et al. (2020) | 2021 | integrate deep learning (LSTM) and blockchain for facilitating a trust based RS | 20.96 | 97.5 |

The decentralized data collection (or knowledge construction) process is highly reliable, since everything is recorded on blockchain, which makes it transparent, auditable, and tamper resistant. Moreover, different works have demonstrated that blockchain-based RSs are able to reach similar accuracy as the centralized schemes, but with much better efficacy by adopting a fully decentralized architecture of the blockchain Casino and Patsakis (2019). Moreover, they enable the preservation of the users' privacy, which is a property that is not provided by traditional centralized

RS engines. However, there are still several limitations and challenges that should be addressed to improve the quality of blockchain-based RSs Harris and Waggoner (2019).

### 3.2. Drawbacks and limitations

Apart from collaborative filtering recommendation algorithms which have been the state-of-the-art RS algorithms for decades, there exist additional approaches, such as knowledge-based and deep learning ones, which have been shown to improve performance and can tackle the cold-start problem of RSs Volkovs et al. (2017); Wei et al. (2017). Their main drawback is that they require that the training data are centrally collected and processed, which limits their applicability on distributed architectures. Apart from a few works, such as Wang et al. (2019a) that take advantage of blockchain and smart contracts in order to promote a collaborative/crowdsourced creation of knowledge for RSs, and DeepChain Weng et al. (2019) that provides a value-driven incentive mechanism based on blockchain in order to allow distributed model training by multiple users, there are not many works that take full advantage of the blockchain capabilities.

Understanding the suitability of blockchain for a task is important for its proper application. RSs perfectly fit this purpose since: i) knowledge is collaboratively contributed by many users who need to trust each other, ii) user data (e.g. ratings or profile information) has to be stored for a long term, if not permanently, and iii) user data must remain immutable even for the administrators of the RS in order to ensure transparency and increase trust.

The latency and scalability issues faced by current blockchain architectures do not yet allow them to scale up to thousands of transactions-per-second rates Kuzlu et al. (2019). This can be an important limitation for web-scale RSs, such as those employed by social networks Eksombatchai et al. (2018); Sharma et al. (2016), which must be able to handle hundreds of millions of users and billion of items and preferences that come as streams, and are requested to provide recommendations in milliseconds.

Moreover, decentralized RSs have been introduced for solving the problems encountered in traditional RSs that have a central authority (which operates as a trusted party with complete control over the RS) via spreading the responsibility and control to the users. However, this can lead to serious concerns when it comes to disputes or misbehaviour. Therefore, the identification of potential unfair exchanges that could occur during the activity between two or different users is of utmost importance in addition to developing adequate solutions that can create fair process, such as the use of atomic swaps inherited from blockchain Lisi et al. (2020).

Another important issue when using blockchain in RSs is related to its energy consumption and hence its environmental implication along with its potential negative impact on climate. Accordingly, the standard process adopted in the blockchain technology for transactions' verification using the proof-of-work process, is "greatly energy hungry" De Vries (2018), because it needs a vast amount of processing power, and thus electricity, for running related computer calculations. In this context, the wide utilization of the blockchain technology can counterbalance climate change mitigation efforts since electricity is still mainly produced using fossil fuels worldwide Sedlmeir et al. (2020).

### 3.3. Open issues

The open issues in the adoption of blockchain in RSs are mainly related to the blockchain itself Casino et al. (2019a). First, the sustainability of the blockchain protocol is challenged by the increased energy consumption of the PoW consensus algorithm, which brings the need for new algorithms. Second, the adoption of blockchain is growing, but there are still businesses that are hesitant, mainly due to the lack of regulation. Third, the various adopters of blockchain have created a wide diversity of implementations that are not inter-operable.

The technological advances in IoT and smartphone devices, have transformed RSs to an integral part of people's life within smart cities or smart home environments. The collected data can range from less sensitive (e.g. implicit or explicit interests and preferences) to very sensitive personal information (e.g. location, vital signs, etc). When data is collected and processed on the cloud, the risk of privacy violation still exists despite anonymisation Sweeney (2000). At the same time, public blockchains still have several limitations concerning privacy and confidentiality [3]. The use of anonymisation or encryption-based mechanisms can be resource demanding and impractical for IoT-based RSs.

Another open issue relates to the dynamic nature of content used by RSs and the ability of an RS to focus on fresh information, while forgetting information that is very old Castells et al. (2015). Although the forgetfulness property is very important in RSs Zou (2019), it is not yet incorporated into the blockchain architecture, allowing outdated transactions to be removed from the ledger. Once this is solved, it will be beneficial for handling scalability issues of blockchain based RSs.

## 4. Future Directions (D)

*D1. Scalability in blockchain-based RSs*

One main challenge of industrial RSs with billions of users and items is scalability Zhao et al. (2019). Both the training and inference parts of the learning process must be effective and efficient respectively. The solutions can be to improve the hardware infrastructure, parallelize or distribute processing, and design more efficient algorithms.

Blockchain may improve the privacy of users and protect from tampering models and data, but this comes at an additional processing cost, which raises scalability issues for the blockchain technology itself. PoW-based blockchains have a limited transaction throughput of few transactions per second, which can be improved when only a fixed set of block validators is employed. Storing models instead of transactions may lower the needs for transaction throughput, and the use of directed acyclic graph (DAG)-based blockchain may remove any upper bounds in the transaction throughput when new transactions arrive fast, which makes it ideal for IoT applications Cao et al. (2020a).

---

[3] https://en.bitcoin.it/wiki/Weaknesses

Sharding, block size increase, and the separation of signatures from transactions are some of the on-chain techniques that have been used for improving the scalability of blockchain. Off-chain solutions include the Lightning network, where only the transaction channel opening is recorded in the main chain and the raiden network, which supports a variety of transaction types apart from blockchain. Finally, child and inter-chain solutions take advantage of additional chains for improving transaction throughput speed Kim et al. (2018); Xie et al. (2019).

Since blockchain-based RSs rely on decentralized architectures, they always have to balance the quality of recommendations with the communication costs. However, the integration of RSs with blockchain can scale more efficiently than traditional RSs, which suffer from the lack of scalability of singular value decomposition (SVD) and other algorithms. Using locality sensitive hashing (LSH) techniques, for example, has been proven beneficial for keeping the item representations (namely signatures) short and allowing fast similarity computation that is inherent for neighborhood-based collaborative filtering Casino and Patsakis (2019).

### D2. Explainability in RSs

Despite the advances in AI and deep learning and the improved performance in many tasks, humans still have the need to interpret and justify the results of any data driven model. Especially in the case of RSs, researchers have focused on the justification of recommendations from very early Symeonidis et al. (2008) and they linked them with the credibility of the system, the user acceptance and trust.

Explainable RSs provide the end users not only with recommendations about items or actions, but also with explanations that clarify *why* these items or actions are recommended Zhang and Chen (2020). Recently, researchers demonstrated experimentally that when the mechanism that triggers recommendations is explained to the end users, this increases the recommendations' acceptance rates Sardianos et al. (2021) and can support behavioral change Hors-Fraile et al. (2019). These recent works employ knowledge and rule-based recommendations, which make it easier to understand and explain how the recommendation mechanism works. In model-based algorithms, such as collaborative filtering, the neighborhood-style explanation is easy and meaningful for the user- and item-based variants (e.g. "similar users to you also like...", "this item is similar to those that you already liked" etc.). However in the case of latent factor models, such as matrix factorization, and deep learning models the same formulation may not convey the reasoning behind the recommendation. Authors in Abdollahi and Nasraoui (2016) used the distribution of ratings within the target user's neighborhood in order to select items that can be recommended with an explanation.

Blockchain can provide to explainable RSs features like transparency of transactions or models employed for the recommendation, immutability of chained blocks, and traceability of non-repudiation. It can also offer smart contracts that allow business logic to be executed automatically and with the consensus of nodes. The future of explainable RSs can be on top of blockchain, in the form of model ensembles, which generate competing recommendations and allow rule-based or knowledge-based models, encoded within the smart contracts to prioritize recommendations and gain user trust, through a holistic approach that guarantees privacy, security and explainability in tandem Nassar et al. (2020).

### D3. Federated Learning

Federated Learning is a research topic that is quickly gaining the interest of the machine learning community, in cases where collaborative training can take place, while preserving the privacy of nodes that contribute their data for the task. In the federated learning scenario multiple parties jointly train a machine learning model, but instead of exchanging their data as a whole, they exchange data summaries or even gradients and models Li et al. (2019).

In the case of RSs, federated learning can help in developing distributed solutions that preserve data privacy and security. As described in Yang et al. (2020) federation in RSs can be horizontal (i.e. items are shared between parties), vertical (i.e. parties share the same user set but different item set or feature spaces) or transfer where neither items nor users are shared among parties. The latter approach takes advantage of the transfer learning approach Pan and Yang (2009), where a model is trained on a party using the local data of this party and then it is reused as the starting point for the model that is trained by the other party.

Blockchain can be used to enhance the resilience of federated learning architectures and avoid tampering of models. The BLADE-FL framework Li et al. (2020a) is a blockchain-based federated learning model, where all clients train their models locally, then encrypt and share them into a pool of models. Only verified models are aggregated in the clients, and the verified models are in a block which is published only when a consensus is reached. New blocks are created and added to the chain at the end of each communication round. The client consensus mechanism makes BLADE-FL tamper-resisted, and the model federation preserves privacy. The solution can be easily adapted to support federated RSs that exchange locally trained models.

### D4. Cooperative Computing in blockchain-based RSs

The term "cooperative computing" was initially coined for highly distributed computing schemes, which employ machine cycles from thousands of computers for solving tasks Borcea et al. (2002); Hegering et al. (1994). Since mobile, edge, and fog computing have become popular, this highly distributed computing paradigm has re-emerged in order to provide more energy efficient solutions Sheng et al. (2015). The blockchain technology requires tremendous computational power, which still makes it impossible for fog networks. However, cooperative computing seems to offer a promising solution that leverages the computational burden. Authors in Wu and Ansari (2020) have introduced a blockchain for fog node clusters, which demonstrates reduced power consumption and need fewer storage space. They employed it to improve security in each cluster through shared (among the cluster nodes) access control lists (ACL). In the proposed implementation older blocks are deleted to keep the chain length constant and the nodes work cooperatively to validate blocks instead of competing like miner nodes do in other blockchains. Similar paradigms are provided in Mendis et al. (2020), which illustrates a decentralized, secure, and privacy-preserving computing paradigm for asynchronous cooperative computing amongst scattered and untrustworthy computing nodes with limited resources, and in Fu et al. (2019) that presents an energy-efficiency aware architecture of cooperative computing (CC) to blockchain-based IoT.

The cluster-based blockchain presented in Wu and Ansari (2020) can be used for scaling up blockchain-based RSs, without inflating the need for resources. Nodes in a local level (e.g. based on their spatial or social proximity) can share preferences and ratings in the blockchain of their cluster and selectively grant or revoke access to other nodes using their common ACL, which is also stored in the cluster blockchain. Nodes that frequently interact with each other will always be in the same cluster, whereas a node that stops interacting with the cluster, will soon loose access to the cluster resources and data. Authors in Liang et al. (2019) describe a similar approach, though without the use of blockchain, for their trust-based recommendation scheme for a vehicular cyber physical systems network. The cooperative computing approach is employed for validating the credibility of user provided data and recommendations.

*D5. Blockchain-based RS on edge/fog devices: Open challenges*

Despite the wide adoption of blockchain in finance and its rapid expansion in many more application domains (e.g., in logistics and healthcare), it is still underutilised in mobile and IoT services. The main reason for this is the limited computational capacity of IoT and mobile devices, in conjunction with the power resources that blockchain tasks may require.

Smart cities, smart homes Damianou et al. (2019) and autonomous vehicles Fu et al. (2020) have already attracted the interest of researchers on IoT and edge or fog computing. Current research focuses on optimising the performance in the core tasks of each domain, and still relies on centralized or cloud based solutions for issues related to privacy, security, robustness and resilience to noise and attacks. Federated learning allows to train more resilient models and protect the privacy and security on the edge.

A combination of off-chain storage of data in P2P distributed systems that allow edge devices to easily connect and disconnect at any time, with a blockchain that stores the hashes of data locations or the trained models, can be a viable solution for edge-based federated learning Zhao et al. (2020). Smart home and consumption monitoring applications can be benefited from the privacy that edge computing for IoT with blockchain can provide Xiong et al. (2018). In this setup, IoT devices can sense, communicate, and exchange information with each other and with the RS, which may run on the cloud. Edge computing devices in radio access networks can provide local computing power, and solve encryption, hashing, and recommendation tasks Sun et al. (2020).

Fog-based RSs Wang et al. (2019b) have to address the information overload issue, which can be performed by abstracting the useful information from the fog environment. In addition, fog-architectures have limited computing and storage capabilities and low service latency. New algorithms need to be designed, that provide recommendations of high quality and in the same time optimize the system performance. The algorithms must balance between sharing and processing (e.g. abstracting) data locally and uploading data to the cloud for further processing.

*D6. Reinforcement learning for blockchain-based RS*

Usually, RSs can be considered as classifiers or predictors in which recommendation task has a sequential nature. In this context, a RS can be represented as a Markov decision process (MDP), where reinforcement learning (RL) algorithms can be deployed for solving it Arulkumaran et al. (2017); Zhao et al. (2018). Typically, the combination of RL and deep learning has led to the paradigm of deep reinforcement learning (DRL), which can be used in the recommendation problem with massive state and action spaces. To that end, an increasing attention is devoted recently to RL-based and DRL-based RSs Huang et al. (2021); Lei et al. (2020). The reader can refer to Gupta and Katarya (2021b) for more details, where the authors have discussed various DRL-based RSs, highlighted their pros and cons and identified their major challenges.

On the other hand, an incentive mechanism plays an essential role for the functionalities of permissionless blockchain-based RSs by incentivizing users for running and securing underlying consensus protocols. Thus, the design of an incentive-compatible mechanism is a great challenge. However, although the advantages of using blockchain in RSs in terms of security and privacy, a large number of public blockchain based RSs still utilize incentive mechanisms that have poorly understandable and largely untested characteristics. To that end, DRL can be deployed for analyzing attacks on incentive mechanisms of blockchain-based RSs. In this regard, SquirRL is proposed in Hou et al. (2019) to automate attack analysis on blockchain incentive mechanisms with DRL. While in Jameel et al. (2020), a RL technique is used for addressing some of a set of challenging problems blockchain-enabled IoT networks, e.g. the transaction throughput enhancement and block time minimization and.

## 5. Conclusion

In this article, we present a comprehensive survey of blockchain-based RSs. We employ a specific taxonomy to to offer a promising solution that leverages the computational burden. Authors in Wu and Ansari (2020) have introduced a blockchain for fog node clusters, which demonstrates reduced power consumption and need fewer storage space. They employed it to improve security in each cluster through shared (among the cluster nodes) access control lists (ACL). In the proposed implementation older blocks are deleted to keep the chain length constant and the nodes work cooperatively to validate blocks instead of competing like miner nodes do in other blockchains. Similar paradigms are provided in Mendis et al. (2020), which illustrates a decentralized, secure, and privacy-preserving computing paradigm for asynchronous cooperative computing amongst scattered and untrustworthy computing nodes with limited resources, and in Fu et al. (2019) that presents an energy-efficiency aware architecture of cooperative computing (CC) to blockchain-based IoT.

The cluster-based blockchain presented in Wu and Ansari (2020) can be used for scaling up blockchain-based RSs, without inflating the need for resources. Nodes in a local level (e.g. based on their spatial or social proximity) can share preferences and ratings in the blockchain of their cluster and selectively grant or revoke access to other nodes using their common ACL, which is also stored in the cluster blockchain. Nodes that frequently interact with each other will always be in the same cluster, whereas a node that stops interacting with the cluster, will soon loose access to the cluster resources and data. Authors in Liang et al. (2019) describe a similar approach, though without the use of blockchain, for their trust-based recommendation scheme for a vehicular cyber physical systems network. The cooperative computing approach is employed for validating the credibility of user provided data and recommendations.

*D5. Blockchain-based RS on edge/fog devices: Open challenges*

Despite the wide adoption of blockchain in finance and its rapid expansion in many more application domains (e.g., in logistics and healthcare), it is still underutilised in mobile and IoT services. The main reason for this is the limited computational capacity of IoT and mobile devices, in conjunction with the power resources that blockchain tasks may require.

Smart cities, smart homes Damianou et al. (2019) and autonomous vehicles Fu et al. (2020) have already attracted the interest of researchers on IoT and edge or fog computing. Current research focuses on optimising the performance in the core tasks of each domain, and still relies on centralized or cloud based solutions for issues related to privacy, security, robustness and resilience to noise and attacks. Federated learning allows to train more resilient models and protect the privacy and security on the edge.

A combination of off-chain storage of data in P2P distributed systems that allow edge devices to easily connect and disconnect at any time, with a blockchain that stores the hashes of data locations or the trained models, can be a viable solution for edge-based federated learning Zhao et al. (2020). Smart home and consumption monitoring applications can be benefited from the privacy that edge computing for IoT with blockchain can provide Xiong et al. (2018). In this setup, IoT devices can sense, communicate, and exchange information with each other and with the RS, which may run on the cloud. Edge computing devices in radio access networks can provide local computing power, and solve encryption, hashing, and recommendation tasks Sun et al. (2020).

Fog-based RSs Wang et al. (2019b) have to address the information overload issue, which can be performed by abstracting the useful information from the fog environment. In addition, fog-architectures have limited computing and storage capabilities and low service latency. New algorithms need to be designed, that provide recommendations of high quality and in the same time optimize the system performance. The algorithms must balance between sharing and processing (e.g. abstracting) data locally and uploading data to the cloud for further processing.

*D6. Reinforcement learning for blockchain-based RS*

Usually, RSs can be considered as classifiers or predictors in which recommendation task has a sequential nature. In this context, a RS can be represented as a Markov decision process (MDP), where reinforcement learning (RL) algorithms can be deployed for solving it Arulkumaran et al. (2017); Zhao et al. (2018). Typically, the combination of RL and deep learning has led to the paradigm of deep reinforcement learning (DRL), which can be used in the recommendation problem with massive state and action spaces. To that end, an increasing attention is devoted recently to RL-based and DRL-based RSs Huang et al. (2021); Lei et al. (2020). The reader can refer to Gupta and Katarya (2021b) for more details, where the authors have discussed various DRL-based RSs, highlighted their pros and cons and identified their major challenges.

On the other hand, an incentive mechanism plays an essential role for the functionalities of permissionless blockchain-based RSs by incentivizing users for running and securing underlying consensus protocols. Thus, the design of an incentive-compatible mechanism is a great challenge. However, although the advantages of using blockchain in RSs in terms of security and privacy, a large number of public blockchain based RSs still utilize incentive mechanisms that have poorly understandable and largely untested characteristics. To that end, DRL can be deployed for analyzing attacks on incentive mechanisms of blockchain-based RSs. In this regard, SquirRL is proposed in Hou et al. (2019) to automate attack analysis on blockchain incentive mechanisms with DRL. While in Jameel et al. (2020), a RL technique is used for addressing some of a set of challenging problems blockchain-enabled IoT networks, e.g. the transaction throughput enhancement and block time minimization and.

## 5. Conclusion

In this article, we present a comprehensive survey of blockchain-based RSs. We employ a specific taxonomy to Blockchain-based Recommender Systemsdescribe existing blockchain-based RS frameworks with reference to different aspects including security and privacy challenges, characteristics and working concept, blockchain potential impact, and types of blockchain-based RSs and applications. We discuss the limitations and open issues before deriving a set of future directions that can help in improving the quality of blockchain-based RSs. Considering the striking features of blockchain, it is possible to implement efficient decentralized RSs with comparable accuracy as centralized RSs, which will preserve data privacy and guarantee security of sensitive information.

Therefore, we focused on outlining how to develop RSs while dealing with users' personal data right and privacy, increasing the security against malicious attacks and increasing the trust of the users in the recommendations engines. Typically, blockchain has been presented as a promising solution that can be embedded in RSs to to bring responses to the open questions, especially those related to security and privacy preservation. Therefore, blockchain can significantly help in improving the recommendation frameworks by ensuring the security, integrity of data, confidentiality and accessibility.

Finally, it is worth noting that developing decentralized blockchain-based RSs can also raise some concerns, e.g. in case of disputes or misbehaviour between users. Therefore, the development of appropriate solutions that can create fair processes should be investigated in the near future, e.g. using atomic swaps inherited from blockchain.

## Acknowledgements

This paper was made possible by National Priorities Research Program (NPRP) grant No. 10-0130-170288 from the Qatar National Research Fund (a member of Qatar Foundation). The statements made herein are solely the responsibility of the authors.## References

, 2018. Toward privacy-preserving personalized recommendation services. Engineering 4, 21–28. doi:https://doi.org/10.1016/j.eng.2018.02.005. cybersecurity.

Abbas, K., Afaq, M., Ahmed Khan, T., Song, W.C., 2020. A blockchain and machine learning-based drug supply chain management and recommendation system for smart pharmaceutical industry. Electronics 9, 852.

Abdollahi, B., Nasraoui, O., 2016. Explainable matrix factorization for collaborative filtering, in: Proceedings of the 25th International Conference Companion on World Wide Web, pp. 5–6.

Adomavicius, G., Bockstedt, J., Curley, S.P., Zhang, J., Ransbotham, S., 2019. The hidden side effects of recommendation systems. MIT Sloan Management Review 60, 1.

Aggarwal, C.C., 2016. Attack-resistant recommender systems, in: Recommender Systems. Springer, pp. 385–410.

Ahn, L.V., Blum, M., Hopper, N.J., Langford, J., 2003. Captcha: Using hard ai problems for security, in: Proceedings of the 22nd International Conference on Theory and Applications of Cryptographic Techniques, Springer-Verlag, Berlin, Heidelberg. p. 294–311.

Al Omar, A., Bosri, R., Rahman, M.S., Begum, N., Bhuiyan, M.Z.A., 2019. Towards privacy-preserving recommender system with blockchains, in: International Conference on Dependability in Sensor, Cloud, and Big Data Systems and Applications, Springer. pp. 106–118.

Ali, O., Ally, M., Dwivedi, Y., et al., 2020. The state of play of blockchain technology in the financial services sector: A systematic literature review. International Journal of Information Management 54, 102199.

Alsalemi, A., Himeur, Y., Bensaali, F., Amira, A., Sardianos, C., Chronis, C., Varlamis, I., Dimitrakopoulos, G., 2021. A micro-moment system for domestic energy efficiency analysis. IEEE Systems Journal 15, 1256–1263. doi:10.1109/JSYST.2020.2997773.

Alsubaei, F., Abuhussein, A., Shiva, S., 2019. Ontology-based security recommendation for the internet of medical things. IEEE Access 7, 48948–48960.

Anandhan, A., Shuib, L., Ismail, M.A., Mujtaba, G., 2018. Social media recommender systems: review and open research issues. IEEE Access 6, 15608–15628.

Andrychowicz, M., Dziembowski, S., Malinowski, D., Mazurek, Ł., 2015. On the malleability of bitcoin transactions, in: International Conference on Financial Cryptography and Data Security, Springer. pp. 1–18.

Arif, Y.M., Nurhayati, H., Kurniawan, F., Nugroho, S.M.S., Hariadi, M., 2020. Blockchain-based data sharing for decentralized tourism destinations recommendation system. International Journal of Intelligent Engineering & System 13, 472–486.

Arora, M., Chopra, A.B., Dixit, V.S., 2020. An approach to secure collaborative recommender system using artificial intelligence, deep learning, and blockchain, in: Intelligent communication, control and devices. Springer, pp. 483–495.

Arulkumaran, K., Deisenroth, M.P., Brundage, M., Bharath, A.A., 2017. Deep reinforcement learning: A brief survey. IEEE Signal Processing Magazine 34, 26–38.

Beg, S., Anjum, A., Ahmad, M., Hussain, S., Ahmad, G., Khan, S., Choo, K.K.R., 2021. A privacy-preserving protocol for continuous and dynamic data collection in iot enabled mobile app recommendation system (mars). Journal of Network and Computer Applications 174, 102874.

Berkovsky, S., Eytani, Y., Kuflik, T., Ricci, F., 2007. Enhancing privacy and preserving accuracy of a distributed collaborative filtering, in: Proceedings of the 2007 ACM conference on Recommender systems, pp. 9–16.

Bhardwaj, R., Datta, D., 2020. Development of a recommender system healthmudra using blockchain for prevention of diabetes. Recommender System with Machine Learning and Artificial Intelligence: Practical Tools and Applications in Medical, Agricultural and Other Industries , 313–327.

Bokhari, S.T., Aftab, T., Nadir, I., Bakhshi, T., 2019. Exploring blockchain-secured data validation in smart meter readings, in: 2019 22nd International Multitopic Conference (INMIC), pp. 1–7. doi:10.1109/INMIC48123.2019.9022772.

Borcea, C., Iyer, D., Kang, P., Saxena, A., Iftode, L., 2002. Cooperative computing for distributed embedded systems, in: Proceedings 22nd International Conference on Distributed Computing Systems, IEEE. pp. 227–236.

Borràs, J., Moreno, A., Valls, A., 2014. Intelligent tourism recommender systems: A survey. Expert Systems with Applications 41, 7370–7389.

Bosri, R., Rahman, M.S., Bhuiyan, M.Z.A., Al Omar, A., 2021. Integrating blockchain with artificial intelligence for privacy-preserving recommender systems. IEEE Transactions on Network Science and Engineering 8, 1009–1018. doi:10.1109/TNSE.2020.3031179.

Broby, D., Paul, G., 2017. The financial auditing of distributed ledgers, blockchain and cryptocurrencies. Journal of Financial Transformation 46, 76–88.

Cai, H., Zhang, F., 2019. An unsupervised method for detecting shilling attacks in recommender systems by mining item relationship and identifying target items. The Computer Journal 62, 579–597.

Cao, B., Zhang, Z., Feng, D., Zhang, S., Zhang, L., Peng, M., Li, Y., 2020a. Performance analysis and comparison of pow, pos and dag based blockchains. Digital Communications and Networks 6, 480–485.

Cao, S., Zhang, G., Liu, P., Zhang, X., Neri, F., 2019. Cloud-assisted secure ehealth systems for tamper-proofing ehr via blockchain. Information Sciences 485, 427–440.Y. Himeur et al.: *Preprint submitted to Elsevier*  Page 21 of 25

Cao, Y., Chen, X., Yao, L., Wang, X., Zhang, W.E., 2020b. Adversarial attacks and detection on reinforcement learning-based interactive recommender systems, in: Proceedings of the 43rd International ACM SIGIR Conference on Research and Development in Information Retrieval, pp. 1669–1672.

Casino, F., Dasaklis, T.K., Patsakis, C., 2019a. A systematic literature review of blockchain-based applications: current status, classification and open issues. Telematics and informatics 36, 55–81.

Casino, F., Patsakis, C., 2019. An efficient blockchain-based privacy-preserving collaborative filtering architecture. IEEE Transactions on Engineering Management 67, 1501–1513.

Casino, F., Patsakis, C., Solanas, A., 2019b. Privacy-preserving collaborative filtering: A new approach based on variable-group-size microaggregation. Electronic Commerce Research and Applications 38, 100895. doi:10.1016/j.elerap.2019.100895.

Castells, P., Hurley, N.J., Vargas, S., 2015. Novelty and diversity in recommender systems, in: Recommender systems handbook. Springer, pp. 881–918.

Chandel, S., Cao, W., Sun, Z., Yang, J., Zhang, B., Ni, T.Y., 2019. A multi-dimensional adversary analysis of rsa and ecc in blockchain encryption, in: Future of Information and Communication Conference, Springer. pp. 988–1003.

Chen, Y., Zhou, M., Zheng, Z., Chen, D., 2019. Time-aware smart object recommendation in social internet of things. IEEE Internet of Things Journal 7, 2014–2027.

Dahihande, J., Jaiswal, A., Pagar, A.A., Thakare, A., Eirinaki, M., Varlamis, I., 2020. Reducing energy waste in households through real-time recommendations, in: Santos, R.L.T., Marinho, L.B., Daly, E.M., Chen, L., Falk, K., Koenigstein, N., de Moura, E.S. (Eds.), RecSys 2020: Fourteenth ACM Conference on Recommender Systems, Virtual Event, Brazil, September 22-26, 2020, ACM. pp. 545–550.

Damianou, A., Angelopoulos, C.M., Katos, V., 2019. An architecture for blockchain over edge-enabled iot for smart circular cities, in: 2019 15th International Conference on Distributed Computing in Sensor Systems (DCOSS), IEEE. pp. 465–472.

De Medio, C., Limongelli, C., Sciarrone, F., Temperini, M., 2020. Moodlerec: A recommendation system for creating courses using the moodle e-learning platform. Computers in Human Behavior 104, 106168.

De Vries, A., 2018. Bitcoin's growing energy problem. Joule 2, 801–805.

Deebak, B., Al-Turjman, F., 2020. A novel community-based trust aware recommender systems for big data cloud service networks. Sustainable Cities and Society 61, 102274.

Deldjoo, Y., DI NOIA, T., MERRA, F.A., 2020a. A survey on adversarial recommender systems: from attack/defense strategies to generative adversarial networks. ACM Comput. Surv. .

Deldjoo, Y., Noia, T.D., Merra, F.A., 2021. A survey on adversarial recommender systems: from attack/defense strategies to generative adversarial networks. ACM Computing Surveys (CSUR) 54, 1–38.

Deldjoo, Y., Schedl, M., Cremonesi, P., Pasi, G., 2020b. Recommender systems leveraging multimedia content. ACM Computing Surveys (CSUR) 53, 1–38.

Deng, X., Huangfu, F., 2019. Collaborative variational deep learning for healthcare recommendation. IEEE Access 7, 55679–55688.

Du, Y., Fang, M., Yi, J., Xu, C., Cheng, J., Tao, D., 2018. Enhancing the robustness of neural collaborative filtering systems under malicious attacks. IEEE Transactions on Multimedia 21, 555–565.

Duriakova, E., Tragos, E.Z., Smyth, B., Hurley, N., Peña, F.J., Symeonidis, P., Geraci, J., Lawlor, A., 2019. Pdmfrec: a decentralised matrix factorisation with tunable user-centric privacy, in: Proceedings of the 13th ACM Conference on Recommender Systems, pp. 457–461.

Egwutuoha, I.P., Levy, D., Selic, B., Chen, S., 2013. A survey of fault tolerance mechanisms and checkpoint/restart implementations for high performance computing systems. The Journal of Supercomputing 65, 1302–1326.

Eirinaki, M., Gao, J., Varlamis, I., Tserpes, K., 2018. Recommender systems for large-scale social networks: A review of challenges and solutions. Future Generation Computer Systems 78, 413–418.

Eirinaki, M., Louta, M.D., Varlamis, I., 2014. A trust-aware system for personalized user recommendations in social networks. IEEE Transactions on Systems, Man, and Cybernetics: Systems 44, 409–421. doi:10.1109/TSMC.2013.2263128.

Eksombatchai, C., Jindal, P., Liu, J.Z., Liu, Y., Sharma, R., Sugnet, C., Ulrich, M., Leskovec, J., 2018. Pixie: A system for recommending 3+ billion items to 200+ million users in real-time, in: Proceedings of the 2018 world wide web conference, pp. 1775–1784.

Fang, M., Yang, G., Gong, N.Z., Liu, J., 2018. Poisoning attacks to graph-based recommender systems, in: Proceedings of the 34th Annual Computer Security Applications Conference, pp. 381–392.

Frey, R., Wörner, D., Ilic, A., 2016a. Collaborative filtering on the blockchain: a secure recommender system for e-commerce. AMCIS 2016 Proceedings , 36.

Frey, R.M., Vuckovac, D., Ilic, A., 2016b. A secure shopping experience based on blockchain and beacon technology, in: Poster Proceedings of the 10th ACM Conference on Recommender Systems (Poster-RecSys 2016), Boston, USA, September 17, 2016, CEUR-WS. p. 03.

Fu, M., Qu, H., Yi, Z., Lu, L., Liu, Y., 2018. A novel deep learning-based collaborative filtering model for recommendation system. IEEE transactions on cybernetics 49, 1084–1096.

Fu, S., Fan, Q., Tang, Y., Zhang, H., Jian, X., Zeng, X., 2019. Cooperative computing in integrated blockchain-based internet of things. IEEE Internet of Things Journal 7, 1603–1612.

Fu, Y., Yu, F.R., Li, C., Luan, T.H., Zhang, Y., 2020. Vehicular blockchain-based collective learning for connected and autonomous vehicles. IEEE Wireless Communications 27, 197–203.

Ghafari, S.M., Beheshti, A., Joshi, A., Paris, C., Mahmood, A., Yakhchi, S., Orgun, M.A., 2020. A survey on trust prediction in online social networks. IEEE Access 8, 144292–144309.

Gillis, N., 2020. Nonnegative Matrix Factorization. SIAM.

Golbeck, J., 2008. Computing with social trust. Springer Science & Business Media.

Gulati, A., Eirinaki, M., 2019. With a little help from my friends (and their friends): Influence neighborhoods for social recommendations, in: The World Wide Web Conference, Association for Computing Machinery, New York, NY, USA. p. 2778–2784.

Guo, Q., Zhuang, F., Qin, C., Zhu, H., Xie, X., Xiong, H., He, Q., 2020. A survey on knowledge graph-based recommender systems. IEEE Transactions on Knowledge and Data Engineering , 1–1doi:10.1109/TKDE.2020.3028705.

Gupta, G., Katarya, R., 2018. A study of recommender systems using markov decision process, in: 2018 Second International Conference on Intelligent Computing and Control Systems (ICICCS), IEEE. pp. 1279–1283.

Gupta, G., Katarya, R., 2019. Recommendation analysis on item-based and user-based collaborative filtering, in: 2019 International Conference on Smart Systems and Inventive Technology (ICSSIT), IEEE. pp. 1–4.

Gupta, G., Katarya, R., 2021a. Enpso: An automl technique for generating ensemble recommender system. Arabian Journal for Science and Engineering , 1–19.

Gupta, G., Katarya, R., 2021b. A study of deep reinforcement learning based recommender systems, in: 2021 2nd International Conference on Secure Cyber Computing and Communications (ICSCCC), IEEE. pp. 218–220.

Harris, J.D., Waggoner, B., 2019. Decentralized and collaborative AI on blockchain, in: 2019 IEEE International Conference on Blockchain (Blockchain), IEEE. pp. 368–375.

Hassan, M.U., Rehmani, M.H., Chen, J., 2019. Deal: Differentially private auction for blockchain-based microgrids energy trading. IEEE Transactions on Services Computing 13, 263–275.

He, P., Yu, G., Zhang, Y., Bao, Y., 2017. Survey on blockchain technology and its application prospects. Computer Science 44, 1–7.

Hegering, H.G., Neumair, B., Gutschmidt, M., 1994. Cooperative computing and integrated system management: A critical comparison of architectural approaches. Journal of Network and Systems Management 2, 283–316.

Himeur, Y., Alsalemi, A., Al-Kababji, A., Bensaali, F., Amira, A., Sardianos, C., Dimitrakopoulos, G., Varlamis, I., 2021. A survey of recommender systems for energy efficiency in buildings: Principles, challenges and prospects. Information Fusion .

Hofmann, F., Wurster, S., Ron, E., Böhmecke-Schwafert, M., 2017. The immutability concept of blockchains and benefits of early standardization, in: 2017 ITU Kaleidoscope: Challenges for a Data-Driven Society (ITU K), IEEE. pp. 1–8.

Hong, M., Jung, J.J., 2021. Multi-criteria tensor model for tourism recommender systems. Expert Systems with Applications 170, 114537.

Hors-Fraile, S., Malwade, S., Luna-Perejon, F., Amaya, C., Civit, A., Schneider, F., Bamidis, P., Syed-Abdul, S., Li, Y.C., De Vries, H., 2019. Opening the black box: Explaining the process of basing a health recommender system on the i-change behavioral change model. IEEE Access 7, 176525–176540.

Hou, C., Zhou, M., Ji, Y., Daian, P., Tramer, F., Fanti, G., Juels, A., 2019. Squirrl: Automating attack analysis on blockchain incentive mechanisms with deep reinforcement learning. arXiv preprint arXiv:1912.01798 .

Hu, R., Guo, Y., Pan, M., Gong, Y., 2019. Targeted poisoning attacks on social recommender systems, in: 2019 IEEE Global Communications Conference (GLOBECOM), IEEE. pp. 1–6.

Huang, L., Fu, M., Li, F., Qu, H., Liu, Y., Chen, W., 2021. A deep reinforcement learning based long-term recommender system. Knowledge-Based Systems 213, 106706.

Hurley, N.J., 2011. Robustness of recommender systems, in: Proceedings of the fifth ACM conference on Recommender systems, pp. 9–10.

Idrees, S.M., Nowostawski, M., Jameel, R., Mourya, A.K., 2021. Security aspects of blockchain technology intended for industrial applications. Electronics 10, 951.

Jamali, M., Ester, M., 2010. A matrix factorization technique with trust propagation for recommendation in social networks, in: Proceedings of the Fourth ACM Conference on Recommender Systems, Association for Computing Machinery, New York, NY, USA. p. 135–142.

Jameel, F., Javaid, U., Khan, W.U., Aman, M.N., Pervaiz, H., Jäntti, R., 2020. Reinforcement learning in blockchain-enabled iiot networks: a survey of recent advances and open challenges. Sustainability 12, 5161.

Jerripothula, K.R., Rai, A., Garg, K., Rautela, Y.S., 2020. Feature-level rating system using customer reviews and review votes. IEEE Transactions on Computational Social Systems 7, 1210–1219.

Jiang, L., Zhang, X., 2019. Bcosn: A blockchain-based decentralized online social network. IEEE Transactions on Computational Social Systems 6, 1454–1466.

Katarya, R., 2018. Reliable recommender system using improved collaborative filtering technique, in: System Reliability Management. CRC Press, pp. 113–119.

Katarya, R., Ranjan, M., Verma, O.P., 2016. Location based recommender system using enhanced random walk model, in: 2016 Fourth International Conference on Parallel, Distributed and Grid Computing (PDGC), IEEE. pp. 33–37.

Katarya, R., Verma, N., 2017. Automatically detection and recommendation in collaborative groups, in: 2017 International Conference on Intelligent Sustainable Systems (ICISS), IEEE. pp. 218–222.

Katarya, R., Verma, O.P., 2018. Recommender system with grey wolf optimizer and fcm. Neural Computing and Applications 30, 1679–1687.

Kaur, H., Kumar, N., Batra, S., 2018. An efficient multi-party scheme for privacy preserving collaborative filtering for healthcare recommender system. Future Generation Computer Systems 86, 297–307.

Khatoon, A., 2020. A blockchain-based smart contract system for healthcare management. Electronics 9, 94.

Kim, S., Kwon, Y., Cho, S., 2018. A survey of scalability solutions on blockchain, in: 2018 International Conference on Information and Communication Technology Convergence (ICTC), IEEE. pp. 1204–1207.

Koren, Y., 2010. Factor in the neighbors: Scalable and accurate collaborative filtering. ACM Trans. Knowl. Discov. Data 4.

Koren, Y., Bell, R., Volinsky, C., 2009. Matrix factorization techniques for recommender systems. Computer 42, 30–37. doi:10.1109/MC.2009.263.

Kula, M., 2015. Metadata embeddings for user and item cold-start recommendations, in: Bogers, T., Koolen, M. (Eds.), Proceedings of the 2nd Workshop on New Trends on Content-Based Recommender Systems co-located with 9th ACM Conference on Recommender Systems (RecSys 2015), Vienna, Austria, September 16-20, 2015., CEUR-WS.org. pp. 14–21.

Kuzlu, M., Pipattanasomporn, M., Gurses, L., Rahman, S., 2019. Performance analysis of a hyperledger fabric blockchain framework: throughput, latency and scalability, in: 2019 IEEE international conference on blockchain (Blockchain), IEEE. pp. 536–540.

Lam, T.Y., Dongol, B., 2020. A blockchain-enabled e-learning platform. Interactive Learning Environments , 1–23.

Lamport, L., Shostak, R., Pease, M., 1982. The byzantine generals problem. ACM Transactions on Programming Languages and Systems 4, 382–401.

Lei, Y., Wang, Z., Li, W., Pei, H., Dai, Q., 2020. Social attentive deep q-networks for recommender systems. IEEE Transactions on Knowledge and Data Engineering , 1–1doi:10.1109/TKDE.2020.3012346.

Li, J., Shao, Y., Ding, M., Ma, C., Wei, K., Han, Z., Poor, H.V., 2020a. Blockchain assisted decentralized federated learning (blade-fl) with lazy clients. arXiv preprint arXiv:2012.02044 .

Li, Q., Wen, Z., Wu, Z., Hu, S., Wang, N., He, B., 2019. A survey on federated learning systems: vision, hype and reality for data privacy and protection. arXiv preprint arXiv:1907.09693 .

Li, X.H., Cao, C.C., Shi, Y., Bai, W., Gao, H., Qiu, L., Wang, C., Gao, Y., Zhang, S., Xue, X., et al., 2020b. A survey of data-driven and knowledge-aware explainable AI. IEEE Transactions on Knowledge and Data Engineering .

Li, Z., Barenji, A.V., Huang, G.Q., 2018. Toward a blockchain cloud manufacturing system as a peer to peer distributed network platform. Robotics and computer-integrated manufacturing 54, 133–144.

Liang, W., Long, J., Weng, T.H., Chen, X., Li, K.C., Zomaya, A.Y., 2019. Tbrs: A trust based recommendation scheme for vehicular cps network. Future Generation Computer Systems 92, 383–398.

Lisi, A., De Salve, A., Mori, P., Ricci, L., 2019. A smart contract based recommender system, in: International Conference on the Economics of Grids, Clouds, Systems, and Services, Springer. pp. 29–42.

Lisi, A., De Salve, A., Mori, P., Ricci, L., 2020. Practical application and evaluation of atomic swaps for blockchain-based recommender systems, in: 2020 the 3rd International Conference on Blockchain Technology and Applications, pp. 67–74.

Lisi, A., De Salve, A., Mori, P., Ricci, L., Fabrizi, S., 2021. Rewarding reviews with tokens: An ethereum-based approach. Future Generation Computer Systems 120, 36–54.

Liu, M., Wu, K., Xu, J.J., 2019. How will blockchain technology impact auditing and accounting: Permissionless versus permissioned blockchain. Current Issues in Auditing 13, A19–A29.

Liu, Z., Li, Z., 2020. A blockchain-based framework of cross-border e-commerce supply chain. International Journal of Information Management 52, 102059.

Lu, Y., 2019. The blockchain: State-of-the-art and research challenges. Journal of Industrial Information Integration 15, 80–90.

Lu, Z., 2018. The architecture of blockchain system across the manufacturing supply chain.

Ma, H., Yang, H., Lyu, M.R., King, I., 2008. Sorec: Social recommendation using probabilistic matrix factorization, in: Proceedings of the 17th ACM Conference on Information and Knowledge Management, Association for Computing Machinery, New York, NY, USA. p. 931–940.

Maatuk, A.M., Elberkawi, E.K., Aljawarneh, S., Rashaideh, H., Alharbi, H., 2021. The covid-19 pandemic and e-learning: Challenges and opportunities from the perspective of students and instructors. Journal of Computing in Higher Education , 1–18.

McKinney, S.A., Landy, R., Wilka, R., 2017. Smart contracts, blockchain, and the next frontier of transactional law. Wash. JL Tech. & Arts 13, 313.

Mehta, B., Hofmann, T., 2008. A survey of attack-resistant collaborative filtering algorithms. IEEE Data Eng. Bull. 31, 14–22.

Mehta, B., Nejdl, W., 2008. Attack resistant collaborative filtering, in: Proceedings of the 31st annual international ACM SIGIR conference on Research and development in information retrieval, pp. 75–82.

Mendis, G.J., Wu, Y., Wei, J., Sabounchi, M., Roche, R., 2020. A blockchain-powered decentralized and secure computing paradigm. IEEE Transactions on Emerging Topics in Computing .

Mikroyannidis, A., Domingue, J., Bachler, M., Quick, K., 2018. Smart blockchain badges for data science education, in: 2018 IEEE Frontiers in Education Conference (FIE), IEEE. pp. 1–5.

Mobasher, B., Burke, R., Bhaumik, R., Williams, C., 2007. Toward trustworthy recommender systems: An analysis of attack models and algorithm robustness. ACM Transactions on Internet Technology (TOIT) 7, 23–es.

Mohammadi, V., Rahmani, A.M., Darwesh, A.M., Sahafi, A., 2019. Trust-based recommendation systems in internet of things: a systematic literature review. Human-centric Computing and Information Sciences 9, 1–61.

Mohan, C., 2019. State of public and private blockchains: Myths and reality, in: Proceedings of the 2019 International Conference on Management of Data, pp. 404–411.

Nassar, M., Salah, K., ur Rehman, M.H., Svetinovic, D., 2020. Blockchain for explainable and trustworthy artificial intelligence. Wiley Interdisciplinary Reviews: Data Mining and Knowledge Discovery 10, e1340.

Nguyen, H.T., Almenningen, T., Havig, M., Schistad, H., Kofod-Petersen, A., Langseth, H., Ramampiaro, H., 2014. Learning to rank for personalised fashion recommender systems via implicit feedback, in: Mining Intelligence and Knowledge Exploration. Springer, pp. 51–61.

Niranjanamurthy, M., Nithya, B., Jagannatha, S., 2019. Analysis of blockchain technology: pros, cons and swot. Cluster Computing 22, 14743–14757.

O'Mahony, M.P., Hurley, N.J., Silvestre, G.C., 2005. Recommender systems: Attack types and strategies, in: AAAI, pp. 334–339.

Pan, S.J., Yang, Q., 2009. A survey on transfer learning. IEEE Transactions on knowledge and data engineering 22, 1345–1359.

Patel, S.B., Bhattacharya, P., Tanwar, S., Kumar, N., 2020. Kirti: A blockchain-based credit recommender system for financial institutions. IEEE Transactions on Network Science and Engineering .

Polatidis, N., Georgiadis, C.K., Pimenidis, E., Stiakakis, E., 2017. Privacy-preserving recommendations in context-aware mobile environments. Information & Computer Security .

Porkodi, S., Kesavaraja, D., 2020. A trust-based recommender system built on iot blockchain network with cognitive framework. Recommender System with Machine Learning and Artificial Intelligence: Practical Tools and Applications in Medical, Agricultural and Other Industries , 291–311.

Pu, P., Chen, L., Hu, R., 2012. Evaluating recommender systems from the user's perspective: survey of the state of the art. User Modeling and User-Adapted Interaction 22, 317–355.

Putra, G.D., Dedeoglu, V., Kanhere, S.S., Jurdak, R., 2020. Towards scalable and trustworthy decentralized collaborative intrusion detection system for iot, in: 2020 IEEE/ACM Fifth International Conference on Internet-of-Things Design and Implementation (IoTDI), IEEE. pp. 256–257.

Quijano-Sánchez, L., Cantador, I., Cortés-Cediel, M.E., Gil, O., 2020. Recommender systems for smart cities. Information systems 92, 101545.

Rahman, A., Nasir, M.K., Rahman, Z., Mosavi, A., Shahab, S., Minaei-Bidgoli, B., 2020. Distblockbuilding: A distributed blockchain-based sdn-iot network for smart building management. IEEE Access 8, 140008–140018.

Resnick, P., Iacovou, N., Suchak, M., Bergstrom, P., Riedl, J., 1994. Grouplens: An open architecture for collaborative filtering of netnews, in: Proceedings of the 1994 ACM Conference on Computer Supported Cooperative Work, Association for Computing Machinery, New York, NY, USA. p. 175–186.

Resnick, P., Sami, R., 2007. The influence limiter: Provably manipulation-resistant recommender systems, in: Proceedings of the 2007 ACM Conference on Recommender Systems, Association for Computing Machinery, New York, NY, USA. p. 25–32. URL: https://doi.org/10.1145/1297231.1297236, doi:10.1145/1297231.1297236.

Resnick, P., Sami, R., 2008. The information cost of manipulation-resistance in recommender systems, in: Proceedings of the 2008 ACM Conference on Recommender Systems, Association for Computing Machinery, New York, NY, USA. p. 147–154. URL: https://doi.org/10.1145/1454008.1454033, doi:10.1145/1454008.1454033.

Rezaimehr, F., Dadkhah, C., 2021. A survey of attack detection approaches in collaborative filtering recommender systems. Artificial Intelligence Review 54, 2011–2066.

Rubio, J.E., Alcaraz, C., Lopez, J., 2017. Recommender system for privacy-preserving solutions in smart metering. Pervasive and Mobile Computing 41, 205–218.

Saha, J., Chowdhury, C., Biswas, S., 2020. Review of machine learning and deep learning based recommender systems for health informatics, in: Deep Learning Techniques for Biomedical and Health Informatics. Springer, pp. 101–126.

Saleem, Y., Rehmani, M.H., Crespi, N., Minerva, R., 2020. Parking recommender system privacy preservation through anonymization and differential privacy. Engineering Reports , e12297.

Sankar, L.S., Sindhu, M., Sethumadhavan, M., 2017. Survey of consensus protocols on blockchain applications, in: 2017 4th International Conference on Advanced Computing and Communication Systems (ICACCS), IEEE. pp. 1–5.

Sardianos, C., Chronis, C., Varlamis, I., Dimitrakopoulos, G., Himeur, Y., Alsalemi, A., Bensaali, F., Amira, A., 2020a. Real-time personalised energy saving recommendations, in: 2020 International Conferences on Internet of Things (iThings) and IEEE Green Computing and Communications (GreenCom) and IEEE Cyber, Physical and Social Computing (CPSCom) and IEEE Smart Data (SmartData) and IEEE Congress on Cybermatics (Cybermatics), IEEE. pp. 366–371.

Sardianos, C., Varlamis, I., Chronis, C., Dimitrakopoulos, G., Alsalemi, A., Himeur, Y., Bensaali, F., Amira, A., 2021. The emergence of explainability of intelligent systems: Delivering explainable and personalized recommendations for energy efficiency. International Journal of Intelligent Systems 36, 656–680.

Sardianos, C., Varlamis, I., Dimitrakopoulos, G., Anagnostopoulos, D., Alsalemi, A., Bensaali, F., Himeur, Y., Amira, A., 2020b. Rehab-c: Recommendations for energy habits change. Future Generation Computer Systems 112, 394–407.

Sardianos, C., Varlamis, I., Eirinaki, M., 2017. Scaling collaborative filtering to large–scale bipartite rating graphs using lenskit and spark, in: 2017 IEEE Third International Conference on Big Data Computing Service and Applications (BigDataService), IEEE. pp. 70–79.

Sarwar, B., Karypis, G., Konstan, J., Riedl, J., 2001. Item-based collaborative filtering recommendation algorithms, in: Proceedings of the 10th International Conference on World Wide Web, Association for Computing Machinery, New York, NY, USA. p. 285–295.

Schafer, J.B., Konstan, J., Riedl, J., 1999. Recommender systems in e-commerce, in: Proceedings of the 1st ACM conference on Electronic commerce, pp. 158–166.

Schuetz, S., Venkatesh, V., 2020. Blockchain, adoption, and financial inclusion in india: Research opportunities. International Journal of Information Management 52, 101936.

Sedlmeir, J., Buhl, H.U., Fridgen, G., Keller, R., 2020. The energy consumption of blockchain technology: beyond myth. Business & Information Systems Engineering 62, 599–608.

Sharma, A., Jiang, J., Bommannavar, P., Larson, B., Lin, J., 2016. Graphjet: Real-time content recommendations at twitter. Proceedings of the VLDB Endowment 9, 1281–1292.

Sheng, Z., Mahapatra, C., Leung, V.C., Chen, M., Sahu, P.K., 2015. Energy efficient cooperative computing in mobile wireless sensor networks. IEEE Transactions on Cloud Computing 6, 114–126.

Stifter, N., Judmayer, A., Weippl, E., 2019. Revisiting practical byzantine fault tolerance through blockchain technologies, in: Security and Quality in Cyber-Physical Systems Engineering. Springer, pp. 471–495.

Sun, C., Li, H., Li, X., Wen, J., Xiong, Q., Zhou, W., 2020. Convergence of recommender systems and edge computing: A comprehensive survey. IEEE Access 8, 47118–47132.

Sweeney, L., 2000. Simple demographics often identify people uniquely. Health (San Francisco) 671, 1–34.

Symeonidis, P., Nanopoulos, A., Manolopoulos, Y., 2008. Providing justifications in recommender systems. IEEE Transactions on Systems, Man, and Cybernetics-Part A: Systems and Humans 38, 1262–1272.

Tarus, J.K., Niu, Z., Kalui, D., 2018. A hybrid recommender system for e-learning based on context awareness and sequential pattern mining. Soft Computing 22, 2449–2461.

Toledo, R.Y., Alzahrani, A.A., Martínez, L., 2019. A food recommender system considering nutritional information and user preferences. IEEE Access 7, 96695–96711.

Tran, T.N.T., Felfernig, A., Trattner, C., Holzinger, A., 2020. Recommender systems in the healthcare domain: state-of-the-art and research issues. Journal of Intelligent Information Systems , 1–31.

Ullah, N., Mugahed Al-Rahmi, W., Alzahrani, A.I., Alfarraj, O., Alblehai, F.M., 2021. Blockchain technology adoption in smart learning environments. Sustainability 13, 1801.

Umekwudo, J.O., Shim, J., 2020. Blockchain technology for mobile applications recommendation systems. Journal of Society for e-Business Studies 24.

Volkovs, M., Yu, G.W., Poutanen, T., 2017. Dropoutnet: Addressing cold start in recommender systems., in: NIPS, pp. 4957–4966.

Wang, S., Huang, C., Li, J., Yuan, Y., Wang, F.Y., 2019a. Decentralized construction of knowledge graphs for deep recommender systems based on blockchain-powered smart contracts. IEEE Access 7, 136951–136961.

Wang, T., Zhang, G., Bhuiyan, M.Z.A., Liu, A., Jia, W., Xie, M., 2018. A novel trust mechanism based on fog computing in sensor–cloud system. Future Generation Computer Systems .

Wang, X., Gu, B., Ren, Y., Ye, W., Yu, S., Xiang, Y., Gao, L., 2019b. A fog-based recommender system. IEEE Internet of Things Journal 7, 1048–1060.

Wei, J., He, J., Chen, K., Zhou, Y., Tang, Z., 2017. Collaborative filtering and deep learning based recommendation system for cold start items. Expert Systems with Applications 69, 29–39.

Wei, P., Xia, S., Chen, R., Qian, J., Li, C., Jiang, X., 2020. A deep-reinforcement-learning-based recommender system for occupant-driven energy optimization in commercial buildings. IEEE Internet of Things Journal 7, 6402–6413.

Weng, J., Weng, J., Zhang, J., Li, M., Zhang, Y., Luo, W., 2019. Deepchain: Auditable and privacy-preserving deep learning with blockchain-based incentive. IEEE Transactions on Dependable and Secure Computing .

Wibowo, F.M., Sidiq, M.F., Akbar, I.A., Basuki, A.I., Rosiyadi, D., 2019. Collaborative whitelist packet filtering driven by smart contract forum, in: 2019 International Seminar on Research of Information Technology and Intelligent Systems (ISRITI), IEEE. pp. 205–210.

Williams, C.A., Mobasher, B., Burke, R., 2007. Defending recommender systems: detection of profile injection attacks. Service Oriented Computing and Applications 1, 157–170.

Wood, G., et al., 2014. Ethereum: A secure decentralised generalised transaction ledger. Ethereum project yellow paper 151, 1–32.

Wu, D., Ansari, N., 2020. A cooperative computing strategy for blockchain-secured fog computing. IEEE Internet of Things Journal 7, 6603–6609.

Wu, D., Lu, J., Zhang, G., 2015. A fuzzy tree matching-based personalized e-learning recommender system. IEEE transactions on fuzzy systems 23, 2412–2426.

Xie, J., Yu, F.R., Huang, T., Xie, R., Liu, J., Liu, Y., 2019. A survey on the scalability of blockchain systems. IEEE Network 33, 166–173.

Xiong, Z., Zhang, Y., Niyato, D., Wang, P., Han, Z., 2018. When mobile blockchain meets edge computing. IEEE Communications Magazine 56, 33–39.

Xu, C., Wang, J., Zhu, L., Zhang, C., Sharif, K., 2019. Ppmr: a privacy-preserving online medical service recommendation scheme in ehealthcare system. IEEE Internet of Things Journal 6, 5665–5673.

Xu, K., Zhang, W., Yan, Z., 2018. A privacy-preserving mobile application recommender system based on trust evaluation. Journal of computational science 26, 87–107.

Yang, L., Tan, B., Zheng, V.W., Chen, K., Yang, Q., 2020. Federated recommendation systems, in: Federated Learning. Springer, pp. 225–239.

Yeh, T.Y., Kashef, R., 2020. Trust-based collaborative filtering recommendation systems on the blockchain. Advances in Internet of Things 10, 37–56.

Yong, B., Shen, J., Liu, X., Li, F., Chen, H., Zhou, Q., 2020. An intelligent blockchain-based system for safe vaccine supply and supervision. International Journal of Information Management 52, 102024.

Zhan, J., Hsieh, C.L., Wang, I.C., Hsu, T.S., Liau, C.J., Wang, D.W., 2010. Privacy-preserving collaborative recommender systems. IEEE Transactions on Systems, Man, and Cybernetics, Part C (Applications and Reviews) 40, 472–476.

Zhang, F., Wang, S., 2020. Detecting group shilling attacks in online recommender systems based on bisecting k-means clustering. IEEE Transactions on Computational Social Systems 7, 1189–1199.

Zhang, F., Zhou, Q., 2014. Hht–svm: An online method for detecting profile injection attacks in collaborative recommender systems. Knowledge-Based Systems 65, 96–105.

Zhang, H., Sun, Y., Zhao, M., Chow, T.W., Wu, Q.J., 2019. Bridging user interest to item content for recommender systems: An optimization model. IEEE transactions on cybernetics 50, 4268–4280.

Zhang, Y., Chen, X., 2020. Explainable recommendation: A survey and new perspectives. Foundations and Trends in Information Retrieval 14, 1–101.

Zhao, X., Xia, L., Zhang, L., Ding, Z., Yin, D., Tang, J., 2018. Deep reinforcement learning for page-wise recommendations, in: Proceedings of the 12th ACM Conference on Recommender Systems, pp. 95–103.

Zhao, Y., Zhao, J., Jiang, L., Tan, R., Niyato, D., Li, Z., Lyu, L., Liu, Y., 2020. Privacy-preserving blockchain-based federated learning for iot devices. IEEE Internet of Things Journal .

Zhao, Z., Hong, L., Wei, L., Chen, J., Nath, A., Andrews, S., Kumthekar, A., Sathiamoorthy, M., Yi, X., Chi, E., 2019. Recommending what video to watch next: a multitask ranking system, in: Proceedings of the 13th ACM Conference on Recommender Systems, pp. 43–51.

Zheng, Z., Xie, S., Dai, H., Chen, X., Wang, H., 2017. An overview of blockchain technology: Architecture, consensus, and future trends, in: 2017 IEEE international congress on big data (BigData congress), IEEE. pp. 557–564.

Zhou, J., Feng, Y., Wang, Z., Guo, D., 2021. Using secure multi-party computation to protect privacy on a permissioned blockchain. Sensors 21, 1540.

Zou, R., 2019. A deep, forgetful novelty-seeking movie recommender model. arXiv preprint arXiv:1909.01811 .